# Planetary model of sunspot emergence: A spectral and autocorrelation analysis.

Ian Edmonds
12 Lentara St, Kenmore, Brisbane, Australia 4069.
Ph/Fax 61 7 3378 6586, ian@solartran.com.au

**Abstract**
This paper is concerned with intermediate range periodicity in the sunspot area spectrum. An empirical model of sunspot area emergence based on Mercury planet conjunctions was developed and the spectra of the model variation and the sunspot area variation compared. By including solar cycle amplitude modulation and the effect of solar magnetic field reversal the model was able to predict fine detail in the sunspot area spectrum. As Mercury planet conjunctions occur predictably it was possible to compare the time variation of band limited components of sunspot area with the corresponding component variations in the model. When the model component variation was stable corresponding components of sunspot area lagged the model variation by a few tens of days. When a 180 degree phase change occurred in the model variation the corresponding component of sunspot area followed the change over an interval of a few hundred days, first by decreasing to zero and then emerging in phase with the model variation. Where periodicities in sunspot area did not match the periods of Fourier components in the model the autocorrelation function of the model variation provided an indirect numerical link to these periodicities. Included in these indirect periodicities were components at periods of about 138, 235, 355, 472, 530, 605, 833, 1190 and 2020 days. Direct periodicities in the model, those linked directly to Fourier components of sunspot area, included components at about 88, 116, 176, 200, 292 and 405 days. The model was developed without allowing for a threshold for sunspot emergence. However, it was found that with a threshold above which the model variation is effective in triggering sunspots, weak Fourier components at the indirect periods were generated. The model with threshold then encompasses all of the significant intermediate range periodicities in sunspot area.

## 1. Introduction and background.

Sunspots emerge on the Sun when tubes of toroidal magnetic flux, at the base of the convection zone ~ 200,000 km below the surface of the Sun, become buoyantly unstable and form loops that float up to the surface to emerge as bipolar concentrations of magnetic flux, Fisher et al (2000). Daily recordings of sunspot area present on the Northern and Southern hemispheres of the Sun were made by the Greenwich Observatory from 1874 to 1976, and from 1976 to the present by the U.S. Air Force. The periodicity in these records most evident and studied is the ~ 11 year solar cycle. However, from the time when Rieger et al (1984) observed a 154 day periodicity in solar flares there has been considerable interest in intermediate range periodicities in sunspot area and in other manifestations of solar activity. With very long records available, e.g. > 50,000 days of recorded sunspot area, high resolution spectra of intermediate range periodicity in the period range 40 to 2000 days is available from methods of spectral analysis such as the Fast Fourier Transform. The intermediate range frequency spectrums of sunspot area North, (SSAN) and sunspot area South, (SSAS), are illustrated in Figure 1 along with the average. When frequency spectra of sunspot data are obtained over the interval of a single solar cycle the spectra, of lower resolution, are less complex and usually show only a few intermediate range periodicities. As a result most studies have been restricted to single solar cycle spectra, (Lean & Brueckner 1989, Lean 1990, Pap et al 1990, Carbonell & Ballester 1990, Carbonell & Ballester 1992, Verma and Joshi 1987, Verma et al 1992, Oliver and Ballester 1995, Oliver et al 1998, Ballester et al 1999, Ballester et al 2004, Krivova & Solanki 2002, Richardson and Cane 2005, Chowdhury et al 2009, Getko 2014, Chowdhury et al 2015, Zaqarashvili et al 2010, Tan and Chen 2013, and Kolotkov et al 2015). However, the periodicities observed differ significantly from one solar cycle to the next. This variability is illustrated in Figure 2 where daily SSAN emergence during two solar cycles, 18 and 19, has been limited to the intermediate range of periodicity, by smoothing the daily data, first by 30 days (S30) then removing the 2000 day average, (S2000). It is evident that periods of about 400 days will feature strongly in a spectral analysis of data in solar cycle 18 while periodicity of about 800 days will feature strongly in a spectrum obtained during solar cycle 19. Due to this variability from solar cycle to solar cycle, intermediate range periodicity in solar activity has been described as intermittent, quasi-periodic and enigmatic.

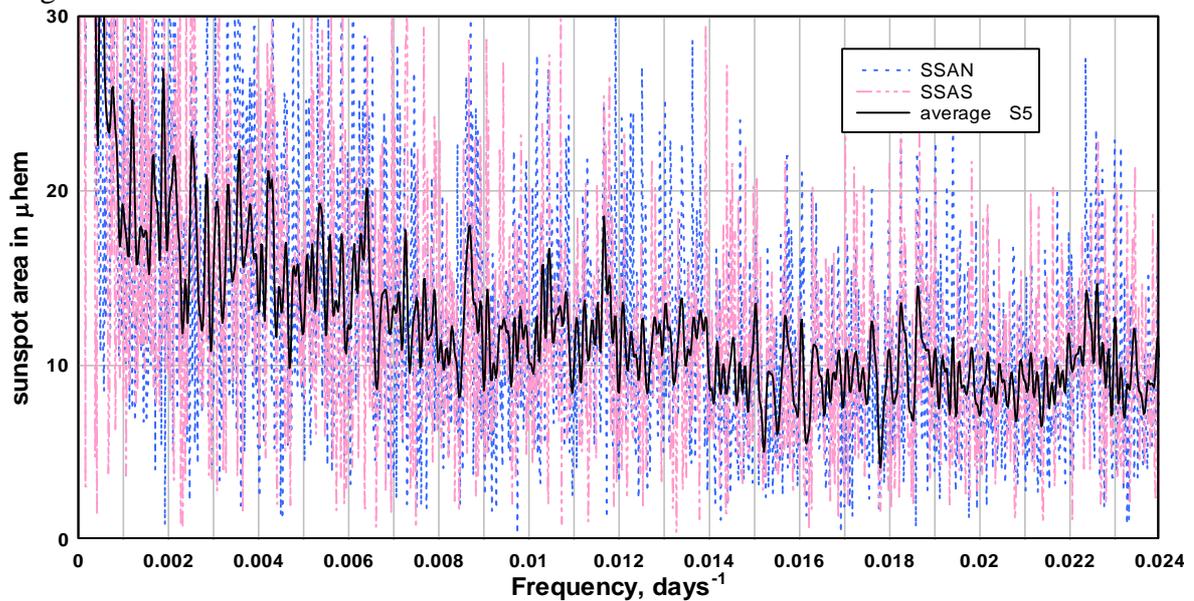

SSAN and SSAS and S3 002 to 012.grf

**Figure 1. The intermediate range frequency spectrums of daily sunspot area North, (SSAN) and sunspot area South, (SSAS), 1876 to 2012. The full line curve is the average spectrum of SSAN and SSAS after a five point smooth.**

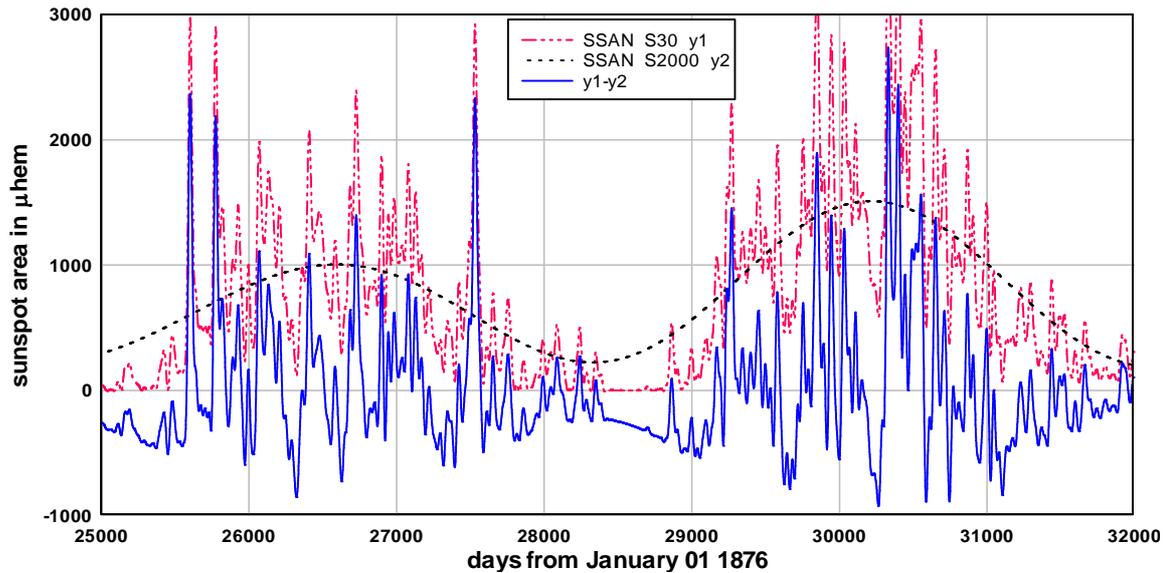

Example SSAN S30 – S2000.grf

**Figure 2.** Illustrates the time variation of SSAN emergence during solar cycles 18 and 19. The periodicity has been limited to the intermediate range by smoothing the daily data, first by 30 days (S30) then removing the 2000 day average, (S2000). Periodicity of about 400 days features strongly in solar cycle 18. Periodicity of about 800 days features strongly during solar cycle 19.

Various explanations for intermediate range periodicity have been advanced, with most suggesting the origin in a periodic source in the Sun, (Bai, 1987, Ichimoto et al, 1985, Bai & Cliver, 1990, Bai & Sturrock, 1991, Sturrock & Bai 1993, Wolff, 1983, 1992, and 1998, Sturrock, 1996, Lockwood, 2001). Wang and Sheeley (2003) demonstrated that intermediate range periodicities in solar magnetic flux can occur through the random development of sunspot groups on the solar surface. Lou (2000), Lou et al (2003), Zaqarashvili et al (2010) and Dimitropolou et al (2008) suggested intermediate range periodicities are linked to equatorially trapped Rossby waves. Seker (2012) suggested periodicities arose from resonance between Alfven waves and planetary tides. Wolff and Patrone (2010) suggested a connection with solar acceleration due to the planets. Other suggestions concerned angular momentum exchanges between planet and Sun, Wilson (2013), and solar inertial motion (Jose 1965, Charvatova 2007). The various proposed planetary mechanisms suffer the defect of producing extremely small effects on the Sun, (De Jager and Versteegh 2005, Callebaut et al 2012, Scafetta 2012). However, helicity oscillations excited by planetary tides may offer an energetically favourable path, Stefani et al (2016). The history of the planetary effects on solar activity has been reviewed by Charbonneau (2002) and a revival of the planetary hypothesis has been discussed by Charbonneau (2013). Currently, there is no generally accepted physical explanation for intermediate range periodicity.

A planetary model of sunspot emergence has the advantage of predictability, in that planet motion is exactly predicable. Thus, in principle, the time dependence of a planetary model should correspond to the time dependence of sunspot emergence. The time dependence of alignment of planets is known exactly and a model based on the time dependence of alignments of several planets could, even in a very simple model, lead to the complexity required to match the spectra and time dependence illustrated in Figures 1 and 2. In this paper we develop a purely empirical planetary model of sunspot emergence, simple enough to have no variable parameters, that allows comparison of the observed time dependence and spectra of sunspot area emergence with the model time dependence and frequency spectrum.

Data and methods are outlined in Section 2. The planetary model is developed in Section 3. Comparison of the spectral detail in the model and in sunspot area is made in Section 4. Section 5 compares the time variation of the model and components of sunspot area emergence during different solar cycles. Section 6 uses the model autocorrelation function to identify components of sunspot area emergence that do not correspond to Fourier components of the model. Section 7 provides evidence of solar magnetic field reversal dependence of sunspot area emergence. Section 8 discusses the correlation observed between North and South hemisphere sunspot emergence. Section 9 is the Conclusion.

## 2. Data and methods

To obtain band pass filtered components of sunspot area a FFT of the entire sunspot area data series was made. The FFTs were obtained with the DPlot application. The resulting n Fourier amplitude and phase pairs, $A(f_n)$, $\phi(f_n)$, in a 20% wide frequency band centred on the component frequency were then used to synthesize the band limited component by summing the n terms, of the form $A(f_n)Cos(2\pi f_n t - \phi(f_n))$ for each day between 1876 and 2012. In the figures and text a band limited component is referred to by the period of the centre of the band. For example, 176SSAN refers to the variation of a component of sunspot area North due to Fourier components in the frequency range $0.00568 +/- 0.00057$ days$^{-1}$, period range 196 to 160 days. Where data has been smoothed the smoothed data is denoted by the suffix Snnn. For example, a 365 day running average of sunspot area North data would be denoted SSAN S365.

In the present study the main variable is the daily sunspot area recorded on the northern hemisphere and southern hemisphere of the Sun (SSAN and SSAS). Sunspot area is measured in units of the area of one millionth of a solar hemisphere or microhems. The data begins in 1874 and is available at http://solarscience.msfc.nasa.gov/greenwch/daily_area.txt. However, due to gaps in the earliest data, we use data from January 01 1876.

The other variable is the orbital radius, $R_M$, of Mercury available at http://omniweb.gsfc.nasa.gov/coho/helios/planet.html for 1959 to 2019. Outside this range past values were calculated using

$R_M = 0.38725 - 0.07975\cos[2\pi t/87.96926 – 1.75]$   AU

where time in days, t, is measured from 0 at January 01, 1995. The planetary data used includes the orbital periods of Mercury, Venus, Earth, Jupiter and Saturn and the relative solar longitude angles of the planets. For example, the relative longitude angle of Mercury and Venus, $\theta_M - \theta_V$, is found from the solar longitudes for Mercury, $\theta_M$, and Venus, $\theta_V$, available at http://omniweb.gsfc.nasa.gov/coho/helios/planet.html .

## 3. A planetary model of periodicity in sunspot area emergence

### 3.1 Development of the model

The concept on which the model is based is that motion of the planets around the Sun induces a periodic time variation of gravitational effect on the Sun that, in turn, triggers the emergence of sunspots with similar periodic variation. There are various types of gravitational effects of planets on the Sun.  Most emphasis has been on studies on tidal effects, e.g. Bigg 1967, Scafetta (2012), Wilson (2013), Seker (2012), Charatova (2007), Hung (2007), Abreu et al (2012), Stefani et al (2016).  In this paper we propose that rapid or pulse like changes in tidal effect are effective in triggering sunspot emergence.  If periodic tidal variations are effective in triggering the emergence of sunspots, similar periodicity should be observed in records of daily sunspot area. We suggest fast or pulse like changes are more effective triggers and focus on the tidal effect associated with Mercury as this planet has the fastest angular speed about the Sun. Mercury also has a strongly elliptical orbit and one form of fast tidal change occurs as Mercury passes through closest approach to the Sun. The tidal effect due to a planet is proportional to $1/R^3$, Scafetta (2012), Svaalgard (2011), Hung (2007). And, due to its highly elliptical orbit, Mercury has the fastest and strongest proportional variation in tidal effect of any planet.  The time variation of $1/R^3$ for Mercury during 1959 is shown in Figure 3. The proportional time variation can be approximated by $1+y_M(t)$ where

$$y_M(t) = 1 + \cos(2\pi t/T_M - \phi) = 1 + \cos(2\pi t/87.969 - 2.2553) \qquad (1)$$

Here $T_M$ is the orbital period of Mercury, 87.969 days, and $\phi$ is the phase of the variation referenced to day t = 0 on January 01, 1876. A scaled version of $1 + y_M(t)$ is shown in Figure 3.

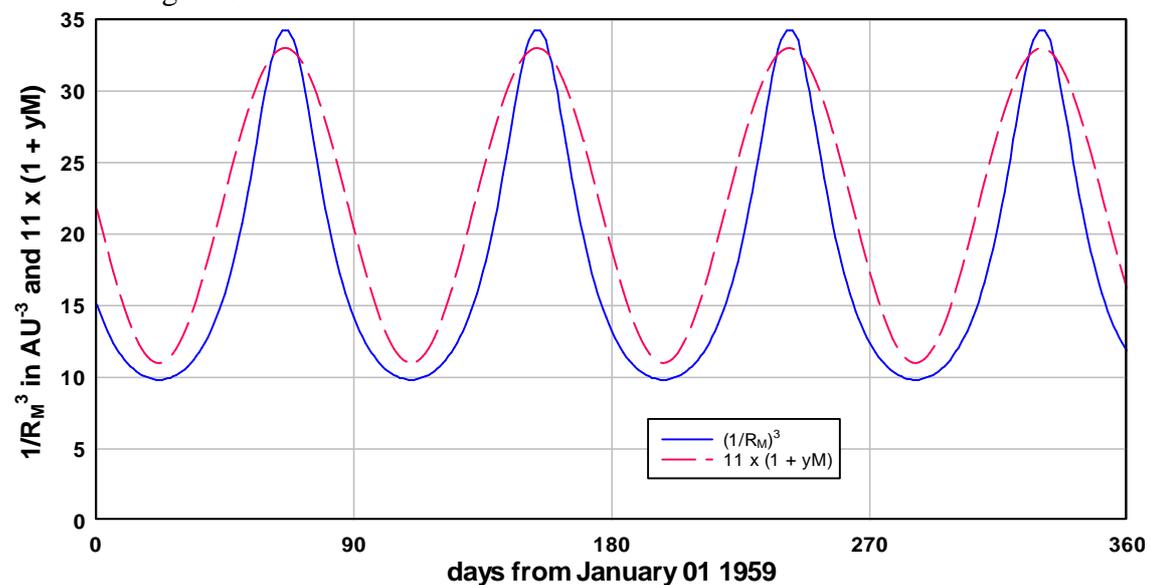

Compare tidal and (1+yM).grf

**Figure 3. The time variation of $1/R_M^3$ for Mercury during 1959 is shown by the full line. This variation can be approximated by $11[1+y_M(t)]$, shown as the broken line, where $y_M(t) = 1 + \cos(2\pi t/87.969 – 2.2553)$. The phase angle of the variation is referenced to t = 0 on January 01, 1876.**

Pulse like changes in tidal effect also occur when Mercury crosses the line connecting the Sun and other planets, i.e. at the time of conjunction. When this occurs there is a pulse in the combined tidal effect as the tide due to Mercury adds to the tide due to the other planet. The conjunction period for Mercury, orbital period $T_M$, and Venus, orbital period $T_V$, is

$T_{MV} = 0.5/(1/T_M – 1/T_V) = 72.2831106$ days.

The conjunction periods of Mercury with the other tidal planets are $T_{ME} = 57.9387398$ days, $T_{MJ} = 44.8961552$ days, $T_{MS} = 44.3473329$ days.

A useful measure of the time variation of the pulse in gravitational effect as Mercury and Venus come into alignment with the Sun can be found from the expression for an alignment index, $I(t) = <\cos(\theta_M(t) – \theta_V(t))>$ where $\theta_M(t)$ and $\theta_V(t)$ are the daily values of the angles of solar longitude of Mercury and Venus, Hung (2007). The time variation of $I(t)$ is not exactly sinusoidal due to the elliptical orbit of Mercury. However, the variation of $I(t)$ can be approximated by $y_{MV}(t)$ where

$$y_{MV}(t) = 1 + \cos(2\pi t/T_{MV} + 1.510) \quad \text{units} \tag{2}$$

and the phase angle is referenced to t = 0 on January 01, 1876. Similarly, the time variation of the tidal pulses due Mercury conjunctions with Earth, Jupiter and Saturn can be approximated by

$$y_{ME}(t) = 1 + \cos(2\pi t/T_{ME} + 0.7744) \quad \text{units} \tag{3}$$

$$y_{MJ}(t) = 1 + \cos(2\pi t/T_M - 3.79115) \quad \text{units} \tag{4}$$

$$y_{MS}(t) = 1 + \cos(2\pi t/T_{MS} - 0.8502) \quad \text{units} \tag{5}$$

The expressions 1 – 5 above are approximations because, due to the elliptical form of the planet orbits, the time variations are not exactly sinusoidal. However, as this paper is primarily concerned with periodicities precise time dependence of the amplitude variation of the tidal effect is not required. It should be mentioned that this paper is not proposing a physical model for the interactions that connect the tidal effect of planets on the Sun to sunspot emergence but is developing an empirical model that can provide a model frequency spectrum and a model time variation that can be correlated with the spectrum and time variation of sunspot emergence. For this reason the time variations of tidal effect, equations 2 to 5 are expressed in arbitrary units.

To ensure a simple model we assume that the amplitude of the effect due to each Mercury- planet conjunction depends only on the amplitude of the Mercury tidal effect illustrated in Figure 3. Thus, in the model, pulses in tidal effect occur at times of Mercury conjunctions with the other planets and the amplitude of the pulses is modulated depending on the proximity of Mercury to the Sun. In this model the time variation of the combined effect is proportional to $z(t)$ where

$$z(t) = [1 + y_M(t)][y_{ME}(t) + y_{MV}(t) + y_{MJ}(t) + y_{MS}(t)] \tag{6}$$

This model has four low frequency components, at the difference frequencies $f_{ME} – f_M$, $f_{MV} - f_M$, $f_{MJ} – f_M$, and $f_{MS} - f_M$. Occasionally two or more of the individual components in $z(t)$ will peak at the same time and larger pulses in $z(t)$ will occur. For example, as Mercury, Earth and Jupiter come into alignment $z(t)$ will have a strong, sharp, pulse. We assume that the triggering of sunspot emergence would depend non-linearly on the amplitude of the pulses, i.e. that larger pulses will be more effective in triggering sunspot emergence than smaller pulses. Modelling non linearity in sunspot emergence could be achieved by setting a threshold level for z(t) above which triggering is supposed to occur, as for example, Scafetta (2012) and Hung (2007). However, this introduces a free variable, other than time, into the model. To keep the model simple we wish to avoid any free variables. We therefore assume that the non linear dependence of sunspot emergence, S(t), is given by a relation of the form

S(t) = Az(t)$^2$                    (7)

where A is a scaling factor. The time variation of z(t)$^2$ is plotted in Figure 4 for the first 2000 days from January 01, 1876. Clearly the contribution to z(t)$^2$ by the four planetary conjunctions yields a complicated time variation comprising pulses of varying amplitude but similar duration, ~ 30 days.

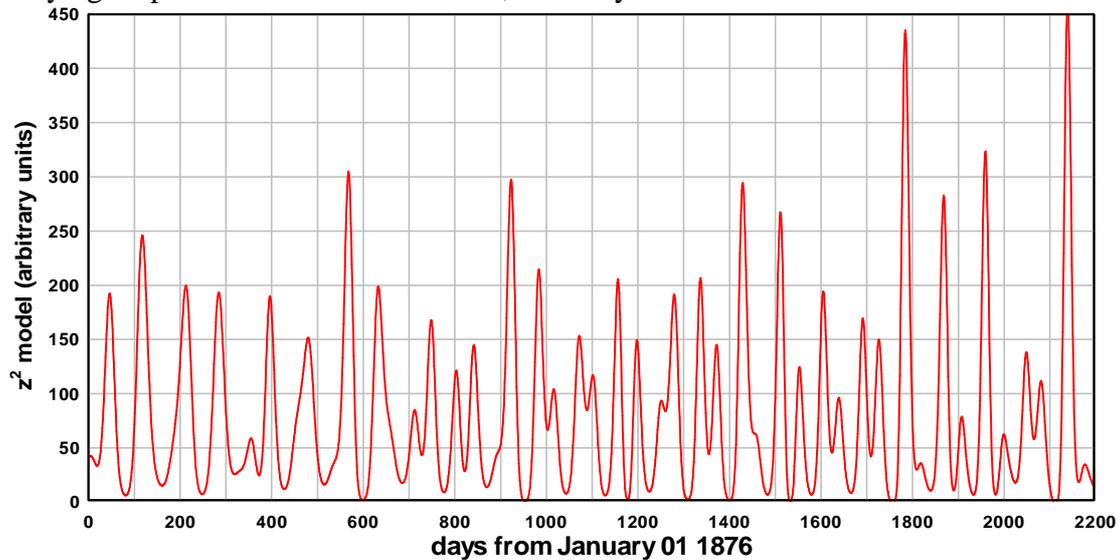

yy and zz models.grf
**Figure 4. The modelled effectiveness for triggering sunspot area emergence, S(t) = z(t)$^2$, for the first 2000 days from January 01, 1876.**

**4. Comparison of the sunspot area spectrum and the model spectrum.**

Expansion of the model time variation into cosine and sine terms yields very complicated expressions and it is much easier to find the frequency of components in the model variation by computing z(t)$^2$ to get the time variation as in Figure 4 and obtaining the Fourier components by Fast Fourier Transform, (FFT). To obtain the high resolution spectrum for comparison with the high resolution sunspot area spectrum z(t)$^2$ is calculated daily from January 01, 1876 to December 31, 2012.

The FFT spectrum of z(t)$^2$ is plotted in Figure 5. We note that the five major spectral peaks occur at the frequencies corresponding to the periods $T_M$, $T_{ME}$, $T_{MV}$, $T_{MJ}$ and $T_{MS}$. The peaks that occur in the intermediate frequency range of interest, 0 to 0.012 days$^{-1}$, are due to components at the frequency differences between the major components. There are also higher frequency components corresponding to sums between the frequencies of the major components that occur in the frequency range around 0.03 days$^{-1}$. The higher frequency components are not of interest in this paper. However, we note that the higher frequency components occur in the spectral range usually associated with the ~27 day rotation period of the Sun.

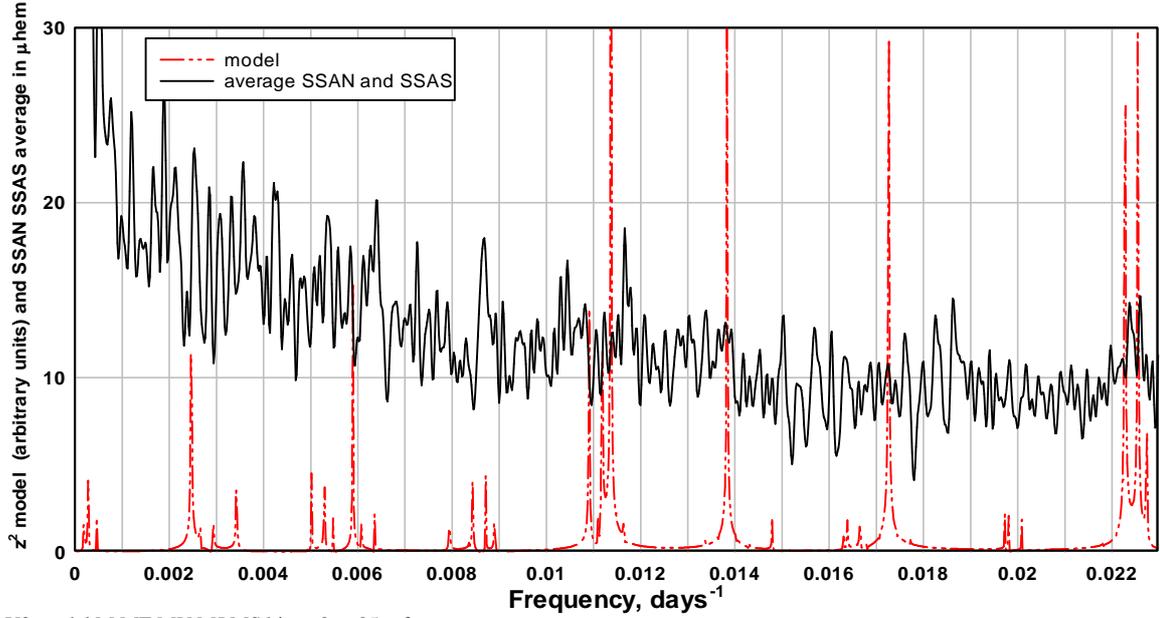

Y2 model M ME MV MJ MS hi res 0 to 05.grf

**Figure 5. The FFT spectrum of $z(t)^2$, the modelled effectiveness of triggering sunspot emergence. The five strongest spectral peaks occur at the frequencies corresponding to the periods $T_M$, $T_{ME}$, $T_{MV}$, $T_{MJ}$ and $T_{MS}$. The weaker peaks that occur in the intermediate frequency range of interest, 0 to 0.012 days$^{-1}$, are due to components at the differences between frequencies of the major components. The full black line is the average spectrum of SSAN and SSAS obtained over the entire record, 1876 to 2012, with 5 point smoothing.**

There are ten strong low frequency difference terms. These occur at frequencies given, for example, by frequency $f = f_M - f_{ME} = 0.00589$ days$^{-1}$ or at period, $T = 1/(1/T_M - 1/T_{ME}) = 170$ days. The frequencies and periods of Fourier components contributing to the model spectrum in the intermediate range are given in Table 1.

**Table 1.**

| Components | Frequency days$^{-1}$ | Period, days | Period, years |
|---|---|---|---|
| $f_{MS}$ | 0.02255 | 44.35 | 0.121 |
| $f_{MJ}$ | 0.02273 | 44.89 | 0.123 |
| $f_{ME}$ | 0.01725 | 57.94 | 0.159 |
| $f_{MV}$ | 0.01383 | 72.28 | 0.198 |
| $f_M$ | 0.01137 | 87.97 | 0.241 |
| $f_{MS} - f_{MJ}$ | 0.00027565 | 3627.8 | 9.93 |
| $f_{MS} - f_{ME}$ | 0.0052897 | 189.04 | 0.517 |
| $f_{MS} - f_{MV}$ | 0.00871478 | 114.747 | 0.314 |
| $f_{MS} - f_M$ | 0.011181 | 89.432 | 0.245 |
| $f_{MJ} - f_{ME}$ | 0.00501401 | 199.441 | 0.546 |
| $f_{MJ} - f_{MV}$ | 0.00843913 | 118.495 | 0.3244 |
| $f_{MJ} - f_M$ | 0.010906 | 91.692 | 0.251 |
| $f_{ME} - f_{MV}$ | 0.0034251 | 291.9607 | 0.7993 |
| $f_{ME} - f_M$ | 0.005892 | 169.721 | 0.4646 |
| $f_{MV} - f_M$ | 0.0024669 | 405.369 | 1.1098 |

Table 1 M ME MV MJ MS.doc

In Figure 5 there are, in the frequency range between 0 and 0.012 days$^{-1}$, five clusters of peaks. When the model spectrum is compared with the average spectrum of SSAN and SSAS, shown as the full line, we note a moderate correspondence between the observed peaks and model peaks. For example at 0.0024 days$^{-1}$ (405 days), at 0.0086 days$^{-1}$ (116 days) and at the pair of peaks corresponding to the frequencies, $f_{MJ}$ = 0.0223 days$^{-1}$ (45 days) and $f_{MS}$ = 0.0225 days$^{-1}$ (44 days). However, it is clear that the sunspot area spectrum contains many more peaks than the spectrum of the model variation provides as it stands. It is well known that sunspot area emergence is strongly amplitude modulated by the ~ 11 year solar cycle. We can include in the model a strong ~ 11 year amplitude modulation by multiplying the z(t) term in equation 6 by the term [1 + cos(2πt/$T_{SC}$)]. $T_{SC}$, the average solar cycle period, is taken as 3982 days, 10.9 years, Scafetta (2012). Including this strong solar cycle modulation in the model results in sidebands at +/- 0.00025 days$^{-1}$ to the model difference frequencies outlined Table 1 and results in twenty additional spectral peaks. For example, in the new model spectrum, Figure 6, the component with the 405 day peak, (frequency 0.00247 days$^{-1}$), now has side band peaks at 0.00222 days$^{-1}$ (450 days) and at 0.00272 days$^{-1}$ (368 days). Similarly, for the other difference frequencies of Table 1 the model will now generate a central peak and two sidebands. However, we note from Figure 6, that the correspondence between observed and predicted peaks is still not very close. Comparison of the sunspot area spectrum and the model spectrum indicates that many of the peaks in the model spectrum coincide, nearly exactly, with minima in the sunspot area spectrum. For example, three of the minima in the sunspot area spectrum are indicated by dotted reference lines in Figure 6. This group of three minima is clearly associated with the $f_{ME}$ - $f_{MV}$ = 0.003425 days$^{-1}$ frequency, 292 day period, component of the model and the two sidebands to the central peak have been generated by the strong ~11 year solar cycle modulation now included in the model. However, the fact that resultant model peaks correspond to minima in the sunspot area spectrum suggests that the sunspot area spectral peaks must be due to a further modulation that results in complete splitting of the original spectral peaks. This further modulation should be accounted for in the model if the model spectrum is to fit the sunspot area spectrum.

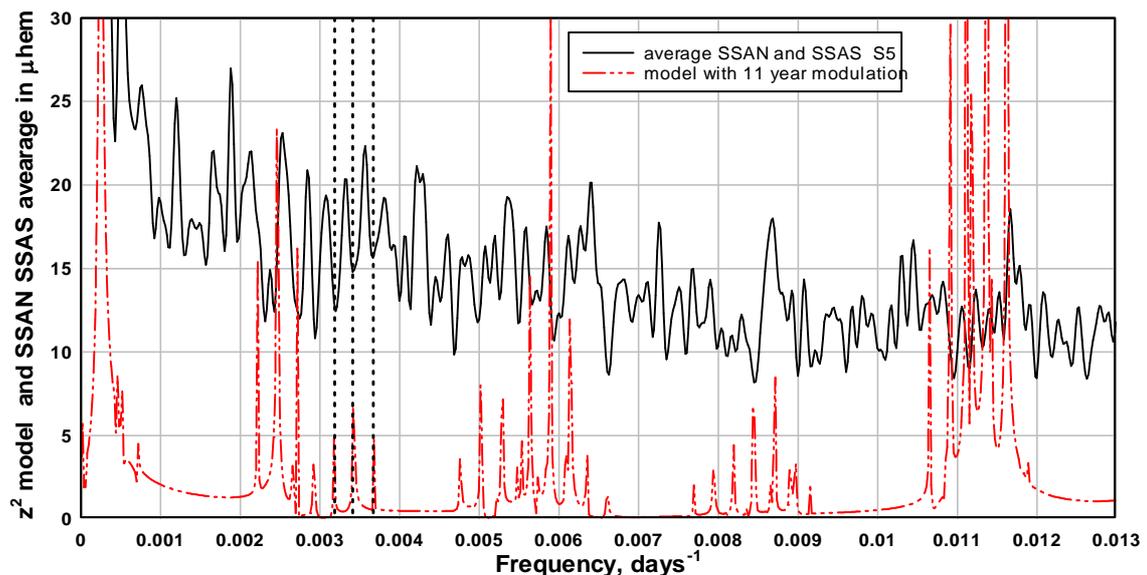

Y2 model M ME MV MJ MS hi res 0 to 013.grf

**Figure 6. When strong ~ 11 year amplitude modulation is included by multiplying the z(t) term in equation 6 by [1 + cos(2πt/$T_{SC}$)] where $T_{SC}$ is the average solar cycle period, the solar cycle modulation results in sidebands at +/- 0.00025 days$^{-1}$ to the model difference frequencies as outlined Table 1 and shown in the spectrum of Figure 5.**

A spectral peak split entirely into two sidebands results when a $\pi$ radian phase shift occurs in the time variation of a component. It is well known that the solar magnetic field reverses around the time of solar cycle maximum. The correspondence of the model spectral peaks with sunspot area spectral minima in Figure 6 suggests that the phase of periodic sunspot emergence associated with the component at the $f_{ME} - f_{MV}$ difference frequency may shift by $\pi$ radians when the polarity of the solar magnetic field reverses. The model outlined above can be modified to take account of a phase reversal during each solar cycle by changing the sign preceding the cosine terms in one or more of the relations of equations 3 to 6. For example, by changing equation 2 from $y_{MV}(t) = 1 + \cos(2\pi t/T_{MV} + 1.510)$ to $y_{MV}(t) = 1 - \cos(2\pi t/T_{MV} + 1.510)$, with the reversal of sign occurring from one solar cycle to the next. This reversal of sign is simply accomplished when computing the model time variation by including the term $\cos(2\pi t/2T_{SC})$ in the computation and changing the sign in the $y_{MV}(t)$ term when the $\cos(2\pi t/2T_{SC})$ term changes sign. Note that the period of this term is ~ 22 years, two times the ~ 11 year solar cycle period, $T_{SC}$. As the effect illustrated in Figure 6 is associated with the component of sunspot emergence due to the $f_{ME} - f_{MV}$ frequency difference, we illustrate the effect of sign reversal on the model spectrum by altering the phase of the component $y_{MV}(t)$ in the model between one solar cycle and the next. The result of this change to the model is illustrated in Figure 7 where model spectrums obtained before and after the change are shown. We note that the three original model peaks associated with the $f_{ME} - f_{MV}$ group are now split by +/- 0.000125 days$^{-1}$ into six peaks. However two of the split peaks overlap with other split peaks and the result is four new peaks in the spectrum. We note that the four model spectrum peaks now align very closely with peaks in the sunspot area spectrum. This suggests that for this component of sunspot area emergence a $\pi$ phase change between one solar cycle and the next occurs frequently. Further support for this observation is provided in section 4 where the time dependence of this component of sunspot area emergence is discussed.

The process just outlined can be summarised as follows. The single peak at 0.003425 days$^{-1}$ is associated with the model component at the $f_{ME} - f_{MV}$ frequency difference. When this component is amplitude modulated by the ~ 11 year solar cycle the resulting spectrum has three peaks, a central peak at 0.003425 days$^{-1}$ and two sideband peaks at 0.003425 +/- 0.00025 days$^{-1}$. When phase reversal of the time variation of the component occurs from solar cycle to solar cycle, phase modulation at ~ 22 year period results in each of the three peaks being split completely into sidebands at +/- 0.000125 days$^{-1}$, resulting in four equally spaced peaks centred on the original frequency difference, $f_{ME} - f_{MV}$. It is evident from the model spectrum in Figure 7 that this type of peak splitting will apply to other components as well and will inevitably lead to complex spectra and broadened spectral peaks in sunspot area, for example the observed broad peak in sunspot area at ~ 0.0086 days$^{-1}$ in Figure 7.

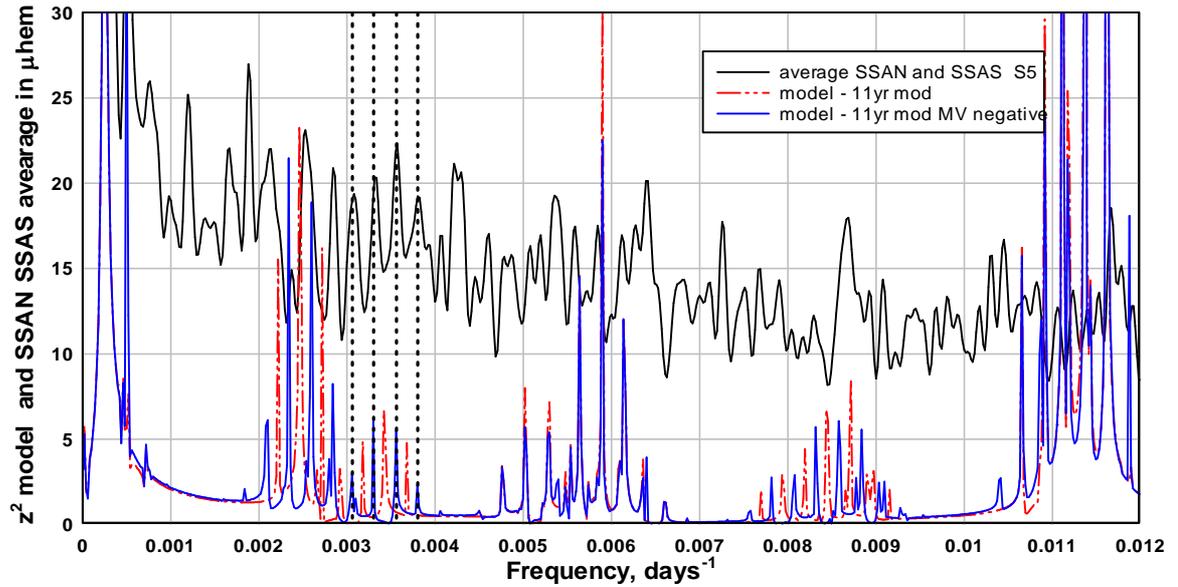

Y2 model M ME MINUS SIGN MV MS.grf

**Figure 7.** Model spectrums obtained before and after changing the sign in the $y_{MV}(t)$ term once every 11 years. We note that the three original model peaks associated with the $f_{ME} - f_{MV}$ group, (red broken line), are now split by +/- 0.000125 days$^{-1}$ to form four new peaks, (blue full line). Note that the four model spectrum peaks now align very closely with the marked peaks in the observed spectrum.

The above analysis indicates that, for significant parts of the sunspot area spectrum, the planetary model as outlined so far can provide a moderately good fit to the observations. However, there are parts of the sunspot area spectrum that do not appear to be associated with the planetary model. For example, the spectral peaks at ~0.0012 days$^{-1}$ (~833 days), ~0.0019 days$^{-1}$ (~526 days), ~0.0042 days$^{-1}$ (~238 days) and ~0.0072 days$^{-1}$ (~139 days) in Figure 7 do not appear to be associated with the model spectrum. We will discuss how these peaks arise from the model in section 5.

One of the simplifying assumptions in deriving equation (6) is that each of the Mercury - planet conjunctions contributes equally to the model variation z(t). For example, the variations $y_{MJ}(t)$ and $y_{MS}(t)$ have been given equal weight in equation (6). Reference to Figure 5 shows that near the frequency 0.0225 days$^{-1}$ there are two equal height peaks in the sunspot area spectrum that correspond closely in frequency to the $f_{MJ}$ and $f_{MS}$ conjunction frequencies, 0.02273 days$^{-1}$ and 0.02255 days$^{-1}$ respectively. This provides observational justification for assigning equal weights to these two components in equation (6) and, by extension, to the other two components.

The analysis of this section indicates that matching a model spectrum to the complex sunspot area spectrum is complicated by the effect of solar cycle modulation and solar magnetic field reversal. In one case, the ~ 290 day component, it was possible to obtain near exact correspondence between the fine detail of the sunspot area spectrum and the fine detail of the model spectrum. In other cases it might be necessary to resort to the even finer detail of the unsmoothed individual spectra of SSAN or SSAS to obtain close fits. To avoid an excessively long paper we change our approach at this point and, in the next section, compare the time dependence of individual components of the model with the time dependence of components of sunspot area during single solar cycles, thus avoiding some of the detail due to solar cycle modulation.

We can summarise the results of this section as follows: A simple planetary model of sunspot emergence can predict accurately the content of some parts of the high resolution frequency spectrum derived from the ~140 year long daily record of sunspot area. The planetary model contains no adjustable parameters other than the possibility of changing by π radians the phase of one or more of the four planetary conjunction components in the model.

**5. Comparing the time dependence of sunspot emergence with the model.**

The planetary model outlined above can, in principle, predict the time of sunspot area emergence. This is a fairly wide assertion. The reason for making it is that the model, other than the possibility of a sign change from solar cycle to solar cycle in the components, contains only one variable, time. If, as seemed to be the case in the previous section, the π phase change happens systematically from one solar cycle to the next, the model would contain no variable other than time and could, in principle, predict sunspot emergence. A factor countering the assertion of predictability is that the model is based on the idea that stronger peaks in the model time variation are more likely to trigger sunspot emergence. Triggering is likely to be probabilistic in function. So, perhaps more accurately, we can state that the model could, in principle, predict the probability of sunspot emergence. The concept of the triggering of sunspot emergence by periodic perturbation of the Sun due to Rossby waves has been discussed, for example, by Lou (2000) and Zaqarashvili et al (2010).

In this section we compare the time variation of components of the model with corresponding components of sunspot emergence. For example, if we are interested in the component of the model associated with the frequency difference $f_{ME} - f_{MV} = 0.003425$ days$^{-1}$, period ~290 days, we generate the model, $S(t) = z(t)^2$ where $z(t) = y_{ME}(t) + y_{MV}(t)$, i.e. we arrange that $z(t)$ contains only the two conjunction terms relevant to this component of sunspot area emergence. We compare the resulting model variation with the component of sunspot area North obtained by filtering the data with a band pass filter centred on the frequency $0.00340$ days$^{-1}$. Throughout this work we use a 20% width frequency band. In the present case the filter band lies between frequencies $0.00340$ +/- $0.00034$ days$^{-1}$, (323 days - 263 days). We refer to the resulting component of sunspot area North as the ~ 290SSAN component. As the model for $z(t)$ does not include the ~11 year modulation we can only apply the comparison between the 290SSAN variation and the model variation one solar cycle at a time. We have not included the ~11 year solar cycle modulation in the model because the time dependence of the solar cycle varies in a more complex manner than the approximation for the solar cycle modulation, $[1 + \cos(2\pi t/T_{SC})]$, used previously. Including the ~ 11 year cycle would necessitate complicating the model.

**5.1 The ~ 290 day periodicity in sunspot area emergence.**
Solar cycle 19 occurs in the interval between day 29000 and day 32000 from January 01, 1876 and is the strongest solar cycle in the record. The model variation, with $z(t) = (y_{ME}(t) + y_{MV}(t))$ and with $y_{MV}(t) = 1 + \cos(2\pi t/T_{MV} + 1.510)$, is computed for t = 29000 to 32000 days and compared with the observed 290SSAN in Figure 8A. A 146 day running average of the model variation is also shown in Figure 8A. It is evident that the model variation is in-phase with the observed 290SSAN component during solar cycle 19. Solar cycle 18 is the second strongest solar cycle. The model variation is computed for t = 25000 to 28000 days and compared with 290SSAN in Figure 8B. It is evident that the 290SSAN variation is in anti-phase with the model variation during this solar cycle. As indicated in the previous section a precise fit to that part of the frequency spectrum associated with this component can be obtained by shifting the phase of the model component by π radians between one solar cycle and the next. When this is done the model variation is brought into phase with the 290SSAN variation in solar cycle 18. However, it is difficult to show this over several solar cycles without an accurate simulation by the model of the actual solar cycle variation.

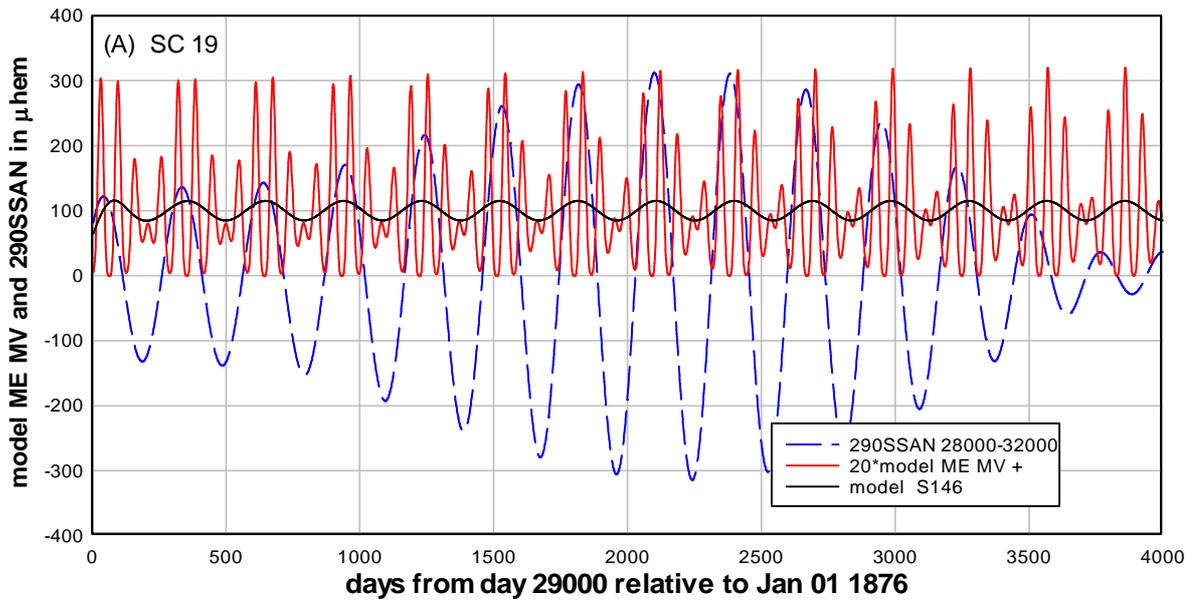

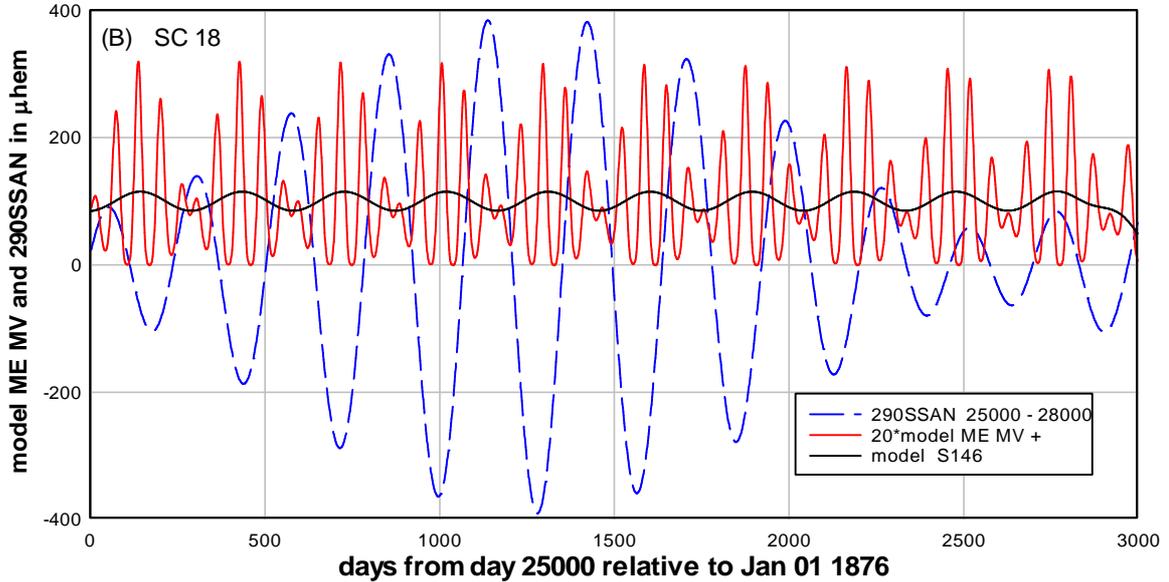

**Figure 8. (A) The variation of 290SSAN during day 29000 to 32000 of solar cycle 19 is compared with the model variation, with z(t) = [y$_{ME}$(t)+ y$_{MV}$(t)] and with y$_{MV}$(t) = 1 + cos(2πt/T$_{MV}$ + 1.510) for the same interval. The 146 day running average of the model variation is also shown. The variations are in-phase. (B) The variation of 290SSAN during day 25000 to 28000 of solar cycle 18 is compared with the model variation, with z(t) = [y$_{ME}$(t)+ y$_{MV}$(t)] and with y$_{MV}$(t) = 1 + cos(2πt/T$_{MV}$ + 1.510), for the same interval. The 146 day running average of the model variation is also shown. The variations are in anti-phase.**

It would be interesting to repeat the comparisons in Figure 8 for all of the twelve solar cycles in the record of sunspot area. Rather than assess all twelve solar cycles in the record one solar cycle at a time, we correlate the 290SSAN variation with a sinusoid of period 291.96 days, the phase angle of which has been adjusted to be in-phase with 290SSAN in solar cycle 19. The result is shown in Figure 9.

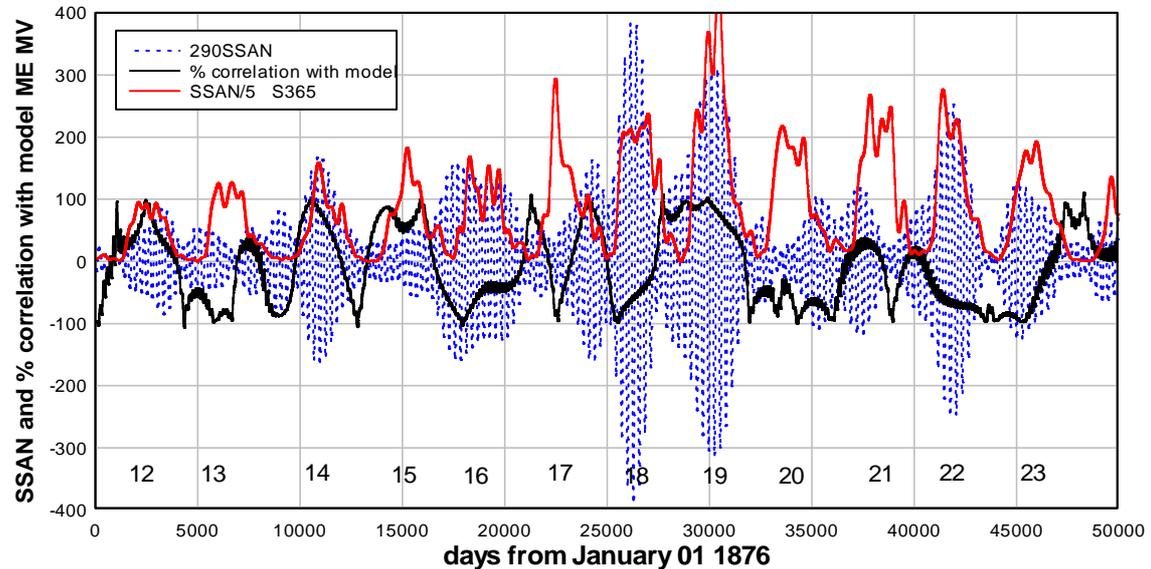

Correlation 290SSAN with ME MV 0-50038.grf
**Figure 9. The entire 290SSAN record, the percentage correlation with a sinusoid of period 292 days and the variation of the SSAN record.**

The percentage correlation between 290SSAN (p) and the sinusoid (q) is found by calculating 100pq/abs(pq) and smoothing the resultant with a 146 day running average. The percentage correlation, in Figure 9, broadly indicates a reversal of phase between one cycle and the next. For example, solar cycles 15, 16, 17, 18, 19, 20, 21 and 22 fit, very approximately, the pattern of phase alternation between one solar cycle and the next. Solar cycles 12, 13 and 14 suggest phase alternation while solar cycle 23 appears to be an exception as its phase is similar to that of solar cycle 22. However, on average over the entire record, a phase alternation pattern between one solar cycle and the next is observed for this component. Phase alternation is consistent with the finding, in the previous section, that reversal of the phase of this component of sunspot emergence, from one solar cycle to the next, is necessary to obtain a fit of the model spectrum to the average SSAN and SSAS spectrum in the frequency band associated with this component. One might expect that the phase alternation would occur at the time of solar magnetic field reversal near the maximum of the solar cycle. However, this does not take into account the possibility of a delayed response, as discussed below. The broader implication of this finding is that, on average, periodic sunspot emergence triggered by the combined effect of Mercury-Earth and Mercury-Venus conjunctions reverses sign as the solar magnetic field changes sign. The physical mechanism by which this might happen is outside the scope of this paper.

**5.2 The ~ 176 day periodicity in sunspot area emergence.**
Figure 6 shows a broad range of model peaks between the frequencies 0.005 days$^{-1}$ and 0.006 days$^{-1}$. The model peaks are associated with the difference frequencies, $f_{ME}$ – $f_M$ = 0.00589 days$^{-1}$ (170 days), $f_{MS}$ - $f_{ME}$ = 0.00529 days$^{-1}$ (189 days) and $f_{MJ}$ - $f_{ME}$

= 0.00501 days$^{-1}$ (199 days). We analyse the relation between the time dependences of the sunspot North data and the model using band pass filtered data obtained using a 20% band pass filter centred on 176 days. The component is referred to as 176SSAN. The entire record of the 176SSAN component is shown in Figure 10. 176SSAN is obviously strongly modulated by the ~11 year solar cycle with mainly one episode of sunspot area emergence in each of solar cycles 19 – 23 and two or three episodes of sunspot area emergence in each of solar cycles 12 to 18.

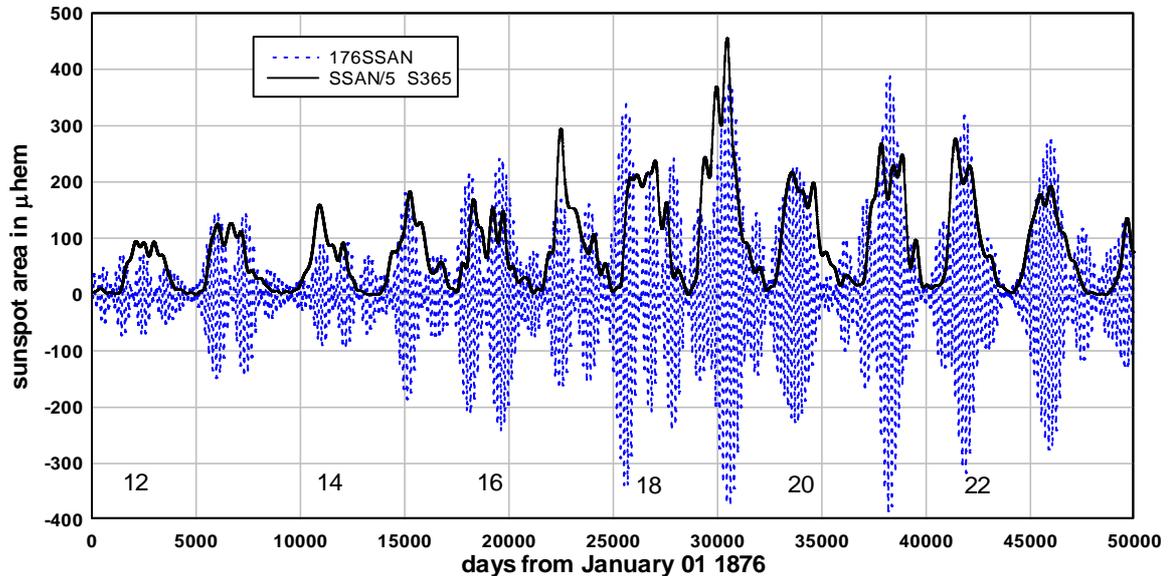

176SSAN entire record.grf

**Figure 10. The entire record of the 176SSAN component is strongly modulated by the ~11 year solar cycle with predominantly one episode of sunspot area emergence in each of solar cycles 19 – 23 and two or three episodes of sunspot area emergence in each of solar cycles 12 to 18.**

The model with $z(t) = (1 + y_M(t))y_{ME}(t)$ is the simplest combination related to the difference frequency $f_{ME} - f_M$ and a periodicity ~170 days, Table 1. We compare the solar cycle 23 variation of model $S(t) = 10z(t)^2$ and the observed variation of 176SSAN in Figure 11. Notice that the model variation is stable from day 0 to day 2000 then undergoes a $\pi$ phase change of the major peaks at about day 2000. Notice that at the beginning of solar cycle 23 the 176SSAN variation moves into near coherency with the model peaks, with the peaks in 176SSAN lagging the model peaks by ~ 20 days. The model peaks change phase at ~ 2000 days into the cycle, i.e. the secondary peak now becomes the stronger peak in the model. The variation in 176SSAN is now out-of-phase with the stronger peak and responds by decreasing in amplitude to about day 2900 when the 176SSAN variation itself undergoes a $\pi$ phase change to bring it into phase with the strongest peak in the model, again developing a lag of ~ 20 days to the model peak. The observations in Figure 11 strongly suggest that this component of sunspot area is being forced by the model variation with a small delay, a few tens of days, when the model variation is stable and responds with a large delay, a few hundred days, when the model variation suffers a $\pi$ phase change. The stable states of the model and the $\pi$ phase changes of the model repeat at intervals of ~2400 days so decreases in amplitude accompanied by a $\pi$ phase change should occur frequently throughout the 50,000 day record.

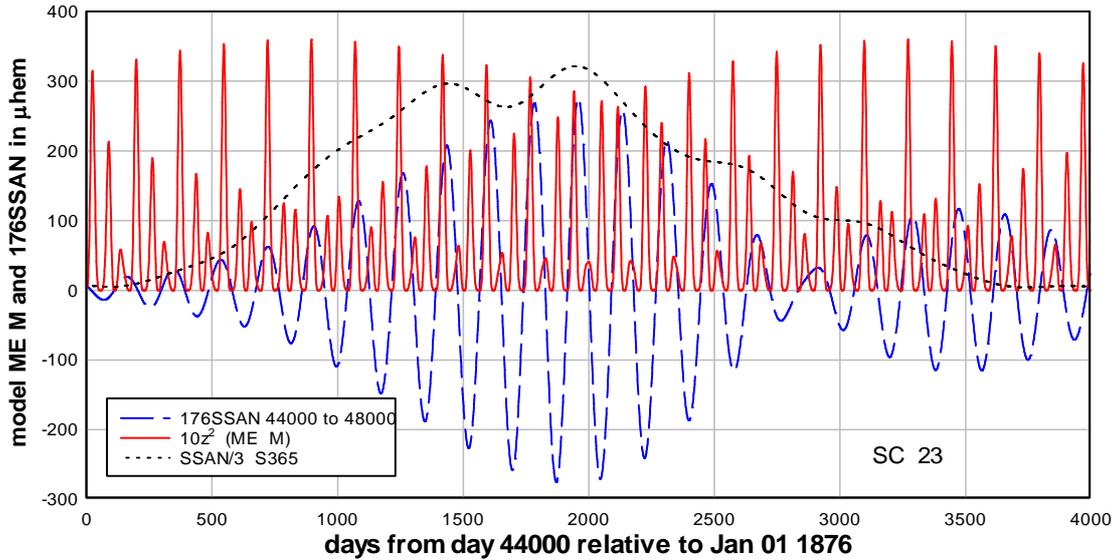

Model M ME and 176SSAN 44000-48000.grf

**Figure 11. At the beginning of solar cycle 23 the 176SSAN variation moves into phase with the model peaks. During the maximum of the cycle the peak in 176SSAN lags the model peak by ~20 days. The model variation changes phase at ~ 2000 days into the cycle, i.e. the secondary peak now becomes the stronger peak in the model. The variation in 176SSAN is now out-of-phase with the stronger peak and decreases sharply in amplitude to about day 2900 when the 176SSAN variation itself undergoes a $\pi$ phase change to bring it into phase with the strongest peak in the model, again developing a lag of about 20 days.**

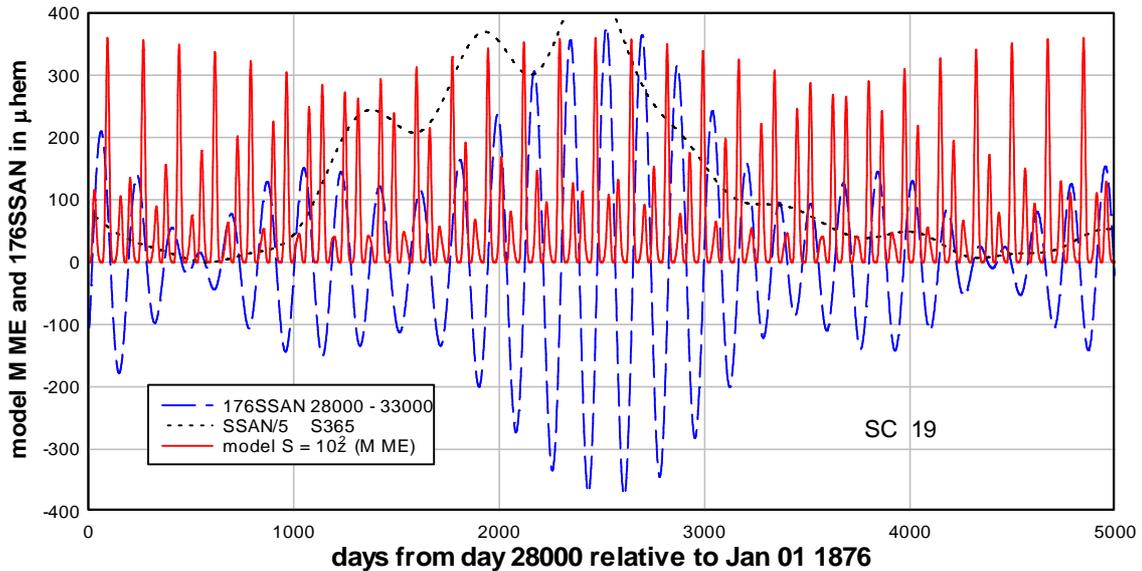

Model M ME and 176SSAN 28000-33000.grf

**Figure 12. Compares the model with the variation of the 176SSAN component in solar cycle 19, the strongest solar cycle in the record. Here we notice that at about day 1000 in Figure 12 the component 176SSAN moves into phase with the increasingly stronger peak of the model. By day 2500 the solar cycle, 176SSAN variation and the model variation are at maximum for this solar cycle and the peaks in 176SSAN lag the peaks in the model by 50 days.**

Figure 12 compares the model with the variation of the 176SSAN component in solar cycle 19, the strongest solar cycle in the record. Here we notice that at the start of the solar cycle, about day 1000 in Figure 12, component 176SSAN moves into phase with the increasingly stronger peak of the model. By day 2500 of the solar cycle, 176SSAN variation and the model variation are at maximum amplitude for this solar cycle and the peaks in 176SSAN lag the peaks in the model by ~ 50 days. The 176SSAN component then decreases as the solar cycle approaches solar minimum. It may be noted that the maximum of solar cycle 19 coincides with a maximum in the stable phase of model peaks. This may account for the strength of the 176SSAN component in solar cycle 19, see Figure 10. We can identify in Figure 12 two times when the major peak in the model changes phase, i.e. times when the secondary peak becomes the major peak. From Figure 12 the times are at ~ 1290 days and at ~ 3660 days. The time difference is ~ 2400 days or ~ 6.5 years. This interval corresponds approximately to the duration of the maximum part of a solar cycle. Therefore this simple form of the model, with $z(t) = (1 + y_M(t))y_{ME}(t)$, should be adequate to follow the variations of this component during solar cycles when there is one episode of sunspot emergence, e.g. solar cycles 19, 20, 21, 22 and 23.

We next consider the earlier solar cycles, 12 – 18, when there was more than one episode of 176SSAN emergence during a solar cycle. The simpler model just discussed, with 6.5 years between the changes in phase of peaks in the model cannot accommodate several episodes of sunspot emergence in any one solar cycle. However, a more complex model, with $z(t) = [1 + y_M(t)][y_{ME}(t) + y_{MJ}(t)]$ can. Figure 13A which compares 176SSAN and the model during solar cycle 17 shows that, in this model the peaks change phase at intervals of 3 - 4 years. Therefore the model variation is able to force two to three episodes of sunspot area emergence during one solar cycle. By smoothing the model variation with an 88 day running average the underlying average variation and periodicity of the model is made more evident. The first two episodes of the model and 176SSAN are separated by ~ 1000 days, (~ 2.7 years) while the second and third episodes of the model and 176SSAN are separated by ~1300 days, (~ 3.5 years). As the time axis is divided into intervals of 176 days it is clear the first phase shift of the model and the first phase shift of the sunspot area component is π radians. The second phase shift of the model at ~2112 days, is 2π radians and the sunspot area component responds slowly to this phase shift in the model, not quite making the full 2π phase shift before the next phase shift of the model. The third model phase shift, this time π radians, occurs at day 3168 after which the sunspot area component returns to lagging the model peaks by ~ 30 days. The slow following of the several model phase changes by the component of sunspot area results in the several episodes of sunspot area emergence during solar cycle 17.

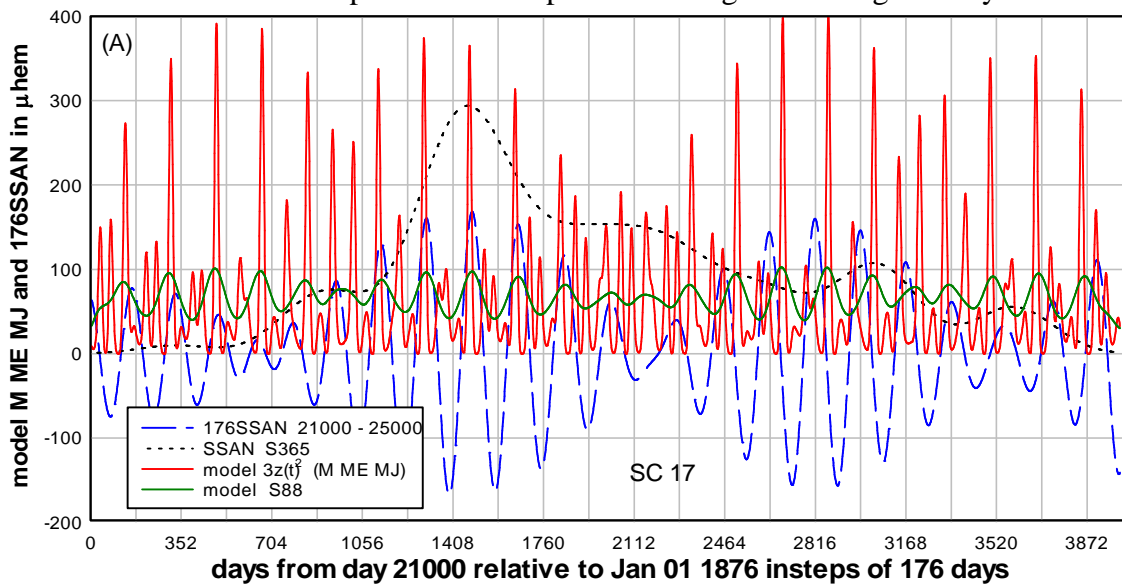

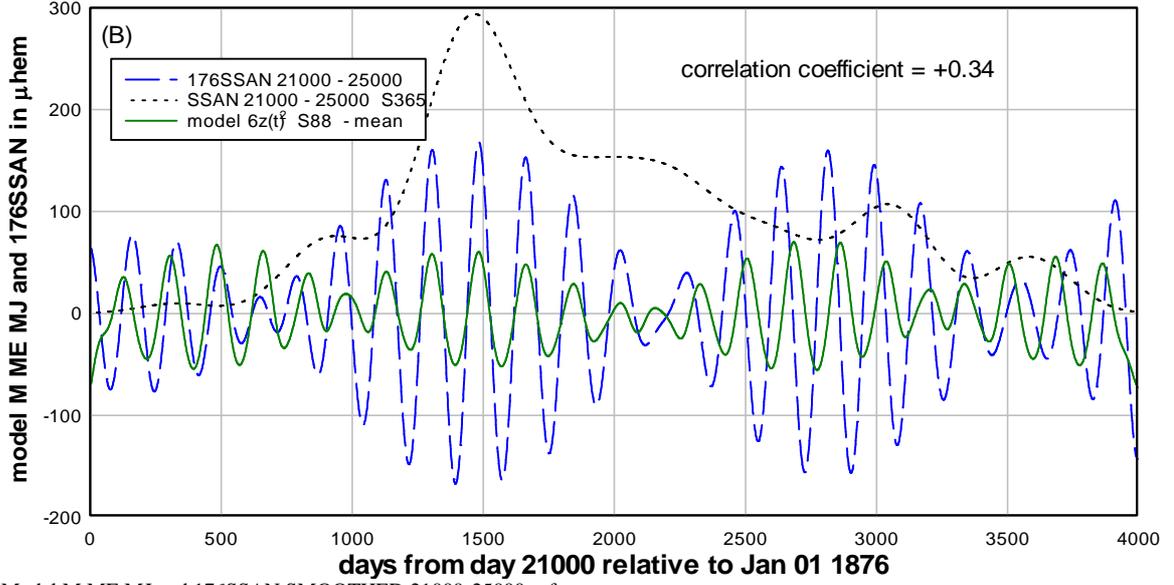

**Figure 13. (A) Compares 176SSAN and the model variations during solar cycle 17. The model variation is able to force two strong episodes of sunspot area emergence during the solar maximum. With smoothing by an 88 day running average the underlying mean periodicity of the model is made more evident. With the time axis is divided into intervals of 176 days it is clear the first phase shift of the model and of the sunspot area component is $\pi$ radians. The second phase shift of the model and the sunspot area component is $2\pi$ radians. (B) Compares the observed 176SSAN and the smoothed version of the model.**

Figure 13B compares the 176SSAN variation and the smoothed variation of the model. The correlation coefficient between the two variations is +0.34. It seems likely that during the earlier solar cycles, 12 – 18, when several episodes of sunspot emergence occurred in each solar cycle, sunspot emergence depended on the $z(t) = [1 + y_M(t)][y_{ME}(t) + y_{MJ}(t)]$ combination. The major components of this combination are at periods ~ 170 days and ~ 200 days, see Table 1. However, there are minor components not listed in Table 1 that also contribute to the model variation, see Figures 6 and 7. As a result the model time variation is more complicated than would be expected of a simple beat between two sinusoidal components. The fact that the 176SSAN variation apparently responds to the model variation, in particular in following the double phase reversal, $2\pi$, as discussed above, provides strong support for the concept of planetary forcing of sunspot emergence.

**5.3 The ~ 116 day periodicity in sunspot area North emergence.**
Figure 6 shows a strong spectral peak at frequency ~ 0.0086 days$^{-1}$ or period ~ 116 days in the average spectrum of sunspot area. This peak is clearly associated with the fairly simple group of peaks in the model associated with the two difference frequencies, $f_{MS} - f_{MV}$ and $f_{MJ} - f_{MV}$, see Table 1. The frequencies are, respectively, 0.008715 days$^{-1}$ and 0.008439 days$^{-1}$, with periods, respectively, 114.7 days and 118.5 days. Because the frequencies are close we expect a long beat period of 3628 days or 9.9 years due to interference of the two components. We expect an average frequency of 0.00858 days$^{-1}$ and the average period in a solar cycle to be ~ 116.6 days. Figure 14A compares 116SSAN during solar cycle 19 with the model $S(t) = 10z(t)^2$ with $z(t) = y_{MV}(t) + y_{MJ}(t)$. The model variation is somewhat complex and it is useful to compare a smoothed version of the model with 116SSAN. This is shown as the green line in Figure 14A. Figure 14B compares 116SSAN during solar cycle 19 with a scaled up version of the smoothed model variation. It is evident that, during solar cycle 19, the ~ 116 day component of sunspot area is coherent with a stable model variation over most of solar cycle 19 consistent with the expectation that there will be, usually, only one episode of this component of sunspot emergence per solar cycle.

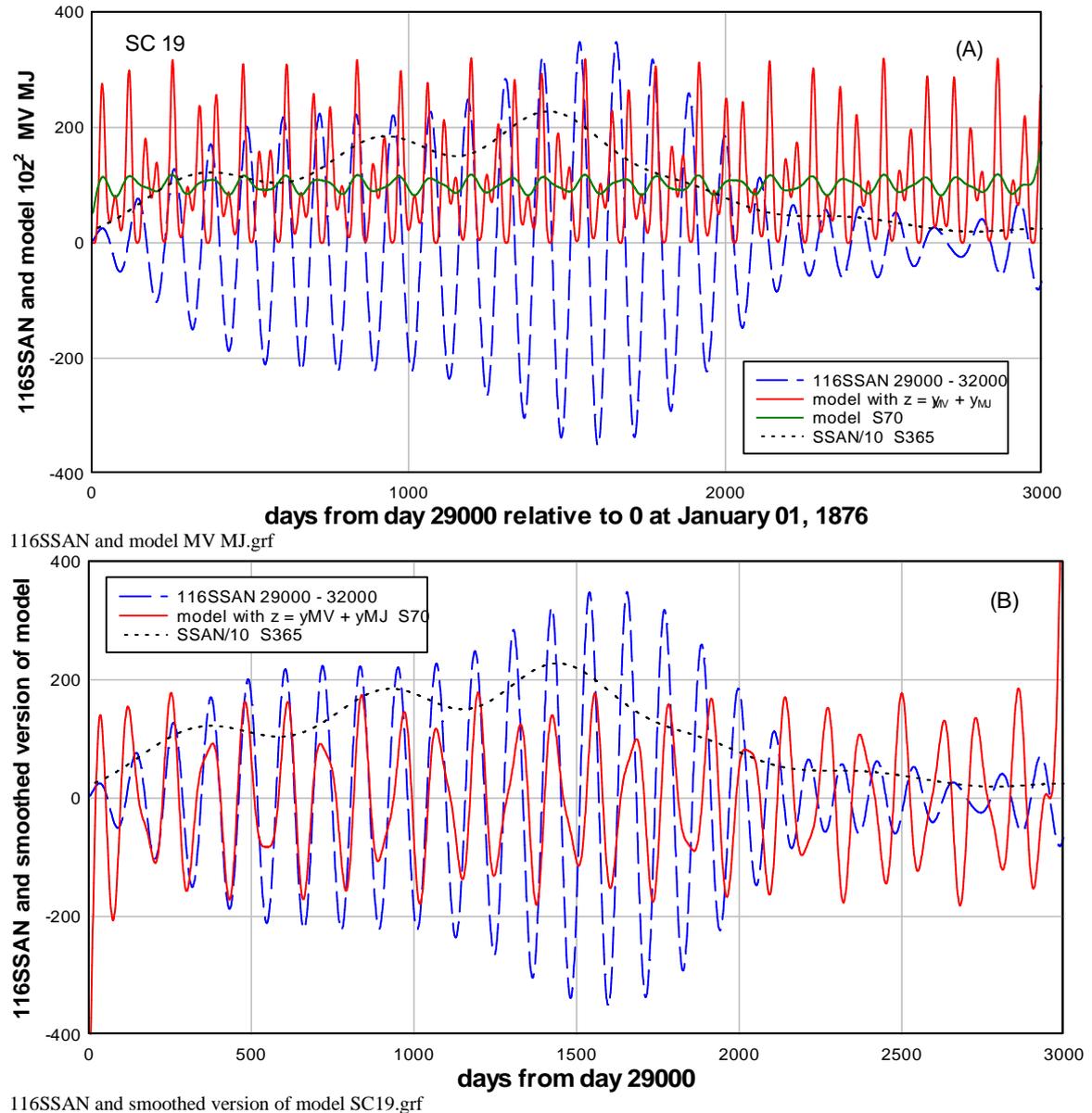

**Figure 14.** (A). Compares 116SSAN during solar cycle 19 with the model $S(t) = 10z(t)^2$ with $z(t) = y_{MV}(t) + y_{MJ}(t)$. The smoothed model variation is also shown. (B) Compares 116SSAN during solar cycle 19 with a scaled up version of the smoothed model variation.

### 5.4 The ~ 88 day periodicity in sunspot area emergence.

The ~ 88 day component variation is difficult to interpret because the relevant difference frequencies are close to the Mercury orbital frequency, see Figure 5. Also, this component is strongly modulated by the ~ 11 year solar cycle and is strongly modulated within a solar cycle, as indicated in Figure 15A. As an example of inter solar cycle modulation Figure 15B shows the variation of the 88SSAN component during solar cycle 23. The four episodes of the component of sunspot emergence evident within solar cycle 23 correspond to some form of fast modulation of the ~ 88 day component in addition to the modulation by the slower ~ 11 year solar cycle. The episodes have the character of a beat pattern due to interference between two equal amplitude components of different frequencies. The model variation $S(t) = z(t)^2$ with $z(t) = y_{MJ}(t)[1 + y_M(t)]$ shown in Figure 15B is dominated by the strong peaks of the $y_M(t)$ variation. The ~ 88 day component of sunspot emergence varies between being in-phase and being out-of-phase with the model peaks as the fast modulation changes and new episodes of sunspot emergence occur. The episodes are separated by about 2 years ~ 730 days, corresponding to a beat pattern due to two near equal amplitude components at $f_1$ and $f_2$ where $f_1 - f_2$ ~ $1/730 = 0.00137$ days$^{-1}$. A frequency difference of this magnitude is available from the peaks at either end of the group of several peaks in the model spectrum near 0.011 days$^{-1}$, Figure 7. However, as the modulation envelope in Figure 15B clearly represents a beat between two near equal amplitude components it is necessary to assess why only two of the widely spaced components of the model within this group are effective during solar cycle 23.

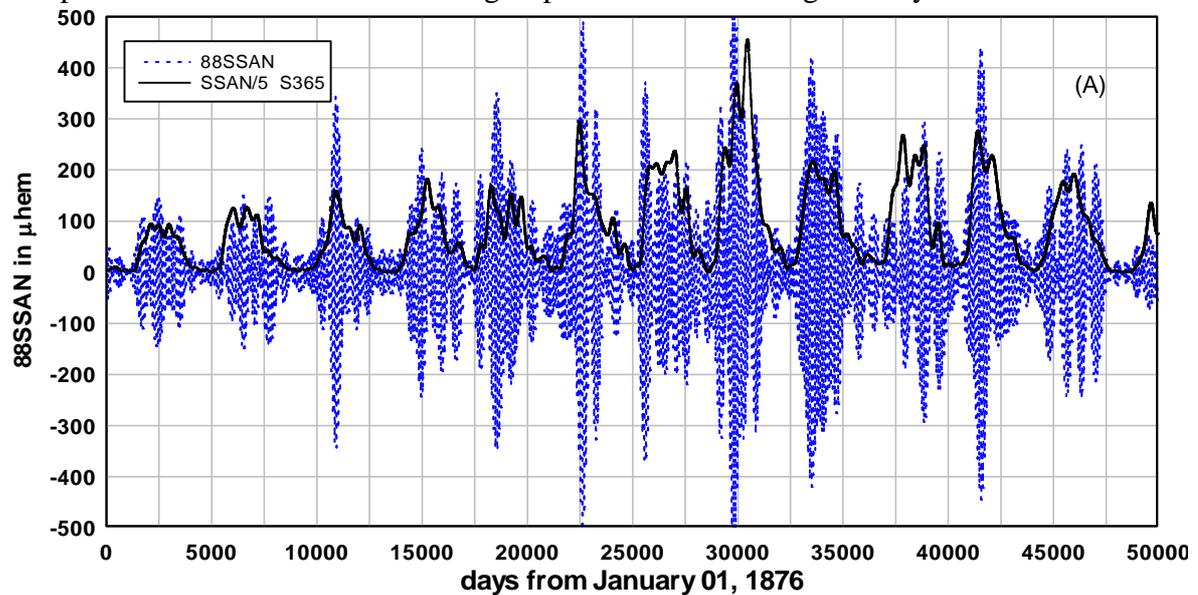

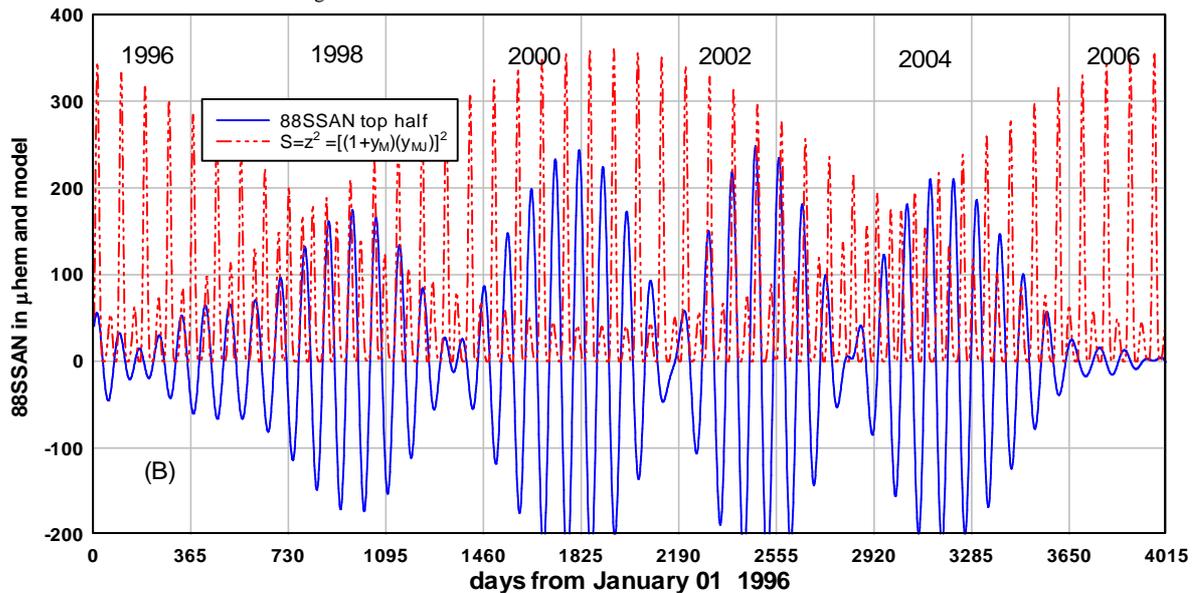

**Figure 15. (A) The ~ 88 day component of SSAN is strongly modulated by the ~ 11 year solar cycle and is strongly modulated within a solar cycle. (B) Shows the variation of the 88SSAN component during solar cycle 23 with four episodes of sunspot emergence evident corresponding**

**to a fast modulation. The fast modulation has the character of a beat pattern due to interference between two equal amplitude components of different frequencies.**

Because the very strong Mercury orbital frequency, $f_M$, occurs within this spectral group it is not as easy to separate the effects due to the frequency difference components, $f_{MS} - f_M = 0.011181$ days$^{-1}$ (89.43 days) and $f_{MJ} - f_M = 0.010906$ days$^{-1}$ (91.69 days) as was the case for more isolated frequency difference groups. For example, the effects due to the isolated $f_{ME} - f_{MV}$ difference frequency at 0.003425 days$^{-1}$ (292 days), Figure 5, were relatively easy to analyse, Figure 6 and Figure 7. In that previous case, see section 3.2, Figure 7, it was possible to follow the splitting of the model peak into peaks at $0.003425 +/- 0.000125$ days$^{-1}$ and $0.003425 +/- 0.000375$ days$^{-1}$ from the effects of solar cycle amplitude modulation and magnetic field reversal phase modulation. It was observed that the resultant four peaks corresponded exactly in frequency to four peaks in the observed average sunspot area spectrum. Assuming that the same type of splitting occurs for the $f_{MJ} - f_M$ component we obtain components at 0.011031, 0.011281, 0.010781 and 0.010531 days$^{-1}$. Similarly the $f_{MS} - f_M$ component splits into components at 0.0110566, 0.011941, 0.0108066 and 0.012191 days$^{-1}$. We observed that, when discussing the splitting of the $f_{ME} - f_{MV}$ frequency component in section 3.2, the resulting components gave rise to near equal amplitude peaks in the sunspot area spectrum. Assuming a similar near equal amplitude response in the present case the resultant sunspot area variation would be a linear combination of eight equal amplitude components at the nominated frequencies above. At present we cannot establish the phase of these components due to the overlap of the strong Mercury orbital component so it is not possible to predict the time variation of the combination. However, by simply adding equal amplitude sinusoidal components at the eight frequencies mentioned above, each with the same phase angle (0), provides some idea of how this type of combination might behave. The combination of the eight sinusoids, computed over 50,000 days, and including modulation by thirteen solar cycles, is shown in Figure 16A. Comparing Figure 16A with Figure 15A it is evident that the combination of eight equal amplitude sinusoids generates the same type of fast modulation pattern evident in the 88SSAN variation. The combination in the solar cycle between 41000 and 45000 days, Figure 16B, shows a similar episode pattern as observed for the 88SSAN component in solar cycle 23, Figure 15B. The result in Figure 16 suggests the following reason for the variability in number and length of episodes of sunspot emergence of the 88SSAN component during a solar cycle. One or two long episodes occur during solar cycles when two or three of the central frequency components of the group, narrowly spaced in frequency, are combining in phase. In solar cycles when the central, narrowly spaced components combine out-of-phase and interfere destructively, the more widely spaced components of the group may interfere to generate three to five episodes. Spectral analysis over the daily variation in solar cycle 23 reveals that the episode pattern of the 88SSAN component in solar cycle 23, Figure 15B is due to the interference of two near equal amplitude components, one at ~ 0.01068 days$^{-1}$ (93.6 days) and the other at ~ 0.01205 days$^{-1}$ (83.0 days), i.e. two components spaced at 0.00137 day$^{-1}$, Edmonds (2016). Similarly a FFT of the model variation in Figure 16B is dominated by two components, one at ~ 0.1050 days$^{-1}$ (95.2 days) and the other at ~ 0.01196 days$^{-1}$ (83.6 days), i.e. due to two components spaced at ~ 0.00146 days$^{-1}$. We note that the frequencies are consistent with the more widely spaced frequencies among the group of eight frequencies estimated above.

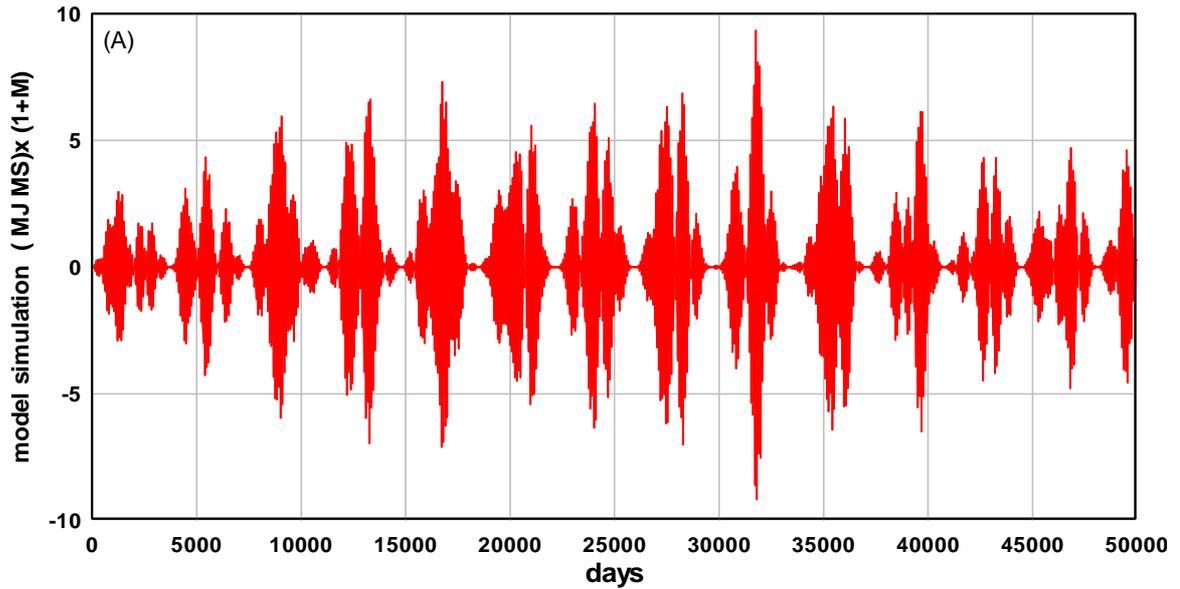

Model simulation MS MJ 1 + M 0 to 50000.grf

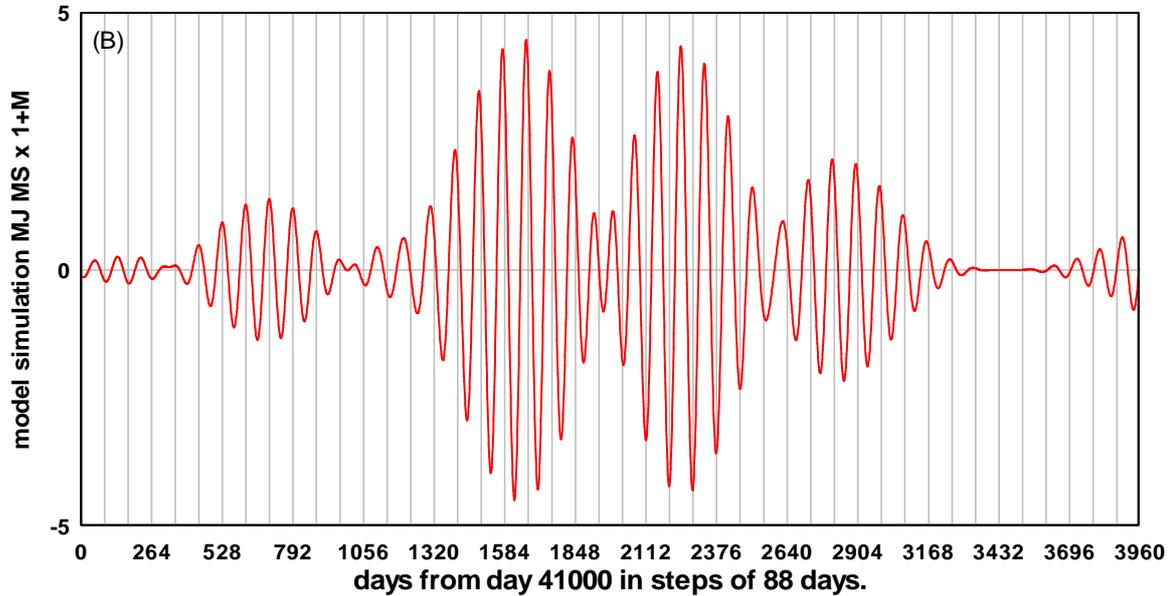

Model simulation MS MJ 1+M 41000.grf

**Figure 16. (A) Shows the combination of eight equal amplitude sinusoids, at the eight frequencies nominated in the text, computed over 50,000 days, and including solar cycle modulation due to thirteen solar cycles. This combination generates the same type of fast modulation pattern evident in the observed 88SSAN variation, see Figure 15A. (B) The episode between 41000 and 45000 days shows a similar modulation pattern as observed for the 88SSAN component in solar cycle 23, Figure 15B. Time axis is in 88 day intervals to facilitate following the phase changes.**

### 5.5 Raw sunspot emergence data compared with band pass filtered data.

The model advanced here is based on the idea that sunspots are more likely to emerge when Mercury is in conjunction with Earth, Venus, Jupiter or Saturn and are more likely to emerge when Mercury is close to the Sun. The combination of these effects results in the model of equations 6 and 7. The time dependent daily variation generated by the model and the time dependent daily variation of sunspot emergence are both very detailed. So for the purpose of comparing the model variation and sunspot area variation we found it useful, in this section, to compare band limited components of the model and sunspot area. For example, in Figure 11 we compared the 176SSAN component with a reduced model $S(t) = z(t)^2$ with $z(t) = [1 +$ $y_M(t)]y_{ME}(t)$, which has a low frequency component at period ~170 days. It is useful to reproduce Figure 11 with the raw daily sunspot data included, Figure 17.

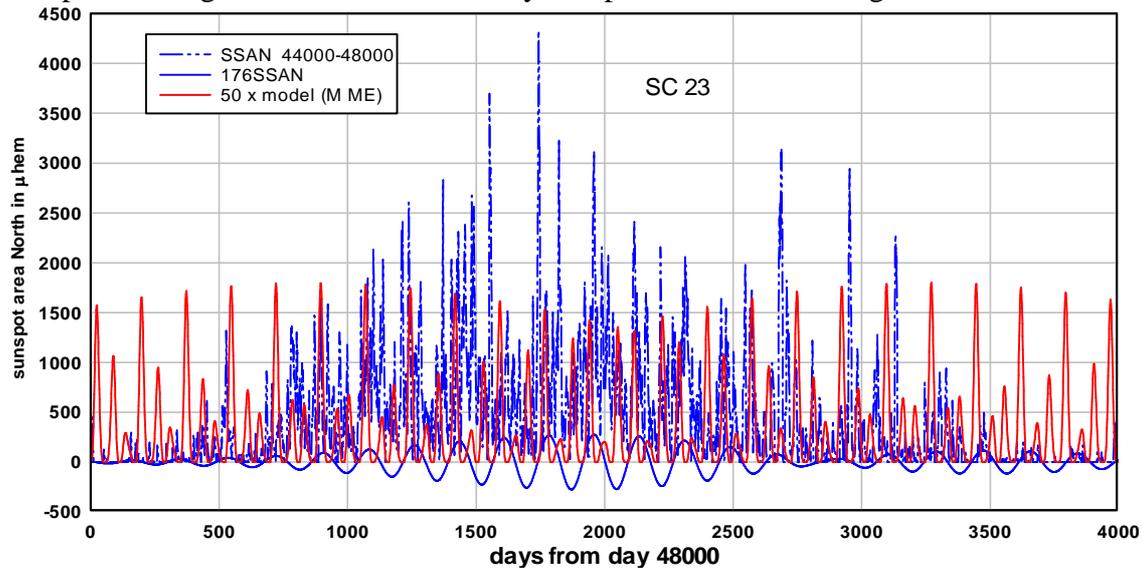

Compares SSAN SC23 with model.grf

**Figure 17. Reproduces Figure 11 with the raw data of SSAN for solar cycle 23 included. While the filtered component, 176SSAN, is consistently coherent with the strong model peaks, the occurrence of the unfiltered daily SSAN is distributed randomly around the model peaks, illustrating the probabilistic nature of the triggering of sunspot emergence.**

Whereas there is close correspondence of the model peaks and the 176SSAN peaks, with the 176SSAN peaks lagging the model peaks by ~ 20 days, the correspondence between the model peaks and the raw sunspot area data is not as clear. Nevertheless, on close examination of Figure 17 it is evident that sunspot area is more likely to emerge near the times when strong model peaks occur. In the case illustrated this corresponds to times when a Mercury-Earth conjunction coincides with closest approach of Mercury to the Sun. This repeats once every three Mercury-Earth conjunctions and once every two Mercury orbits. The comparison with the raw data as in Figure 17 makes it clear that the relationship between the model variation and sunspot emergence is essentially probabilistic, i.e. sunspots are more likely to emerge near the times when major peaks in the model occur.

It is worth noting that preferred times for sunspot emergence lead to a possible explanation of the preferred longitudes or "active longitudes" for sunspot emergence observed, for example, by Berdyugina and Usoskin (2003). The principal component of the planetary model outlined here is the ~ 88 day periodicity due to stronger model peaks occurring at closest approach of Mercury to the Sun. Near solar cycle maximum sunspots tend to emerge on the Sun at about +/- 20$^o$ solar latitude. At these latitudes the solar surface rotates through 360$^o$ of longitude every ~ 25 days. With sunspots more likely to emerge at intervals of 88 days it is easy to show that this leads to the occurrence of preferred longitudes for sunspot emergence, as illustrated in Figure 18. In Figure 18 the preferred longitudes migrate slowly with time. When the solar surface rotation period is 25.15 days the migration rate is zero and the preferred longitudes remain constant with Mercury, at closest approach above the same point on the solar surface every seven solar rotations.

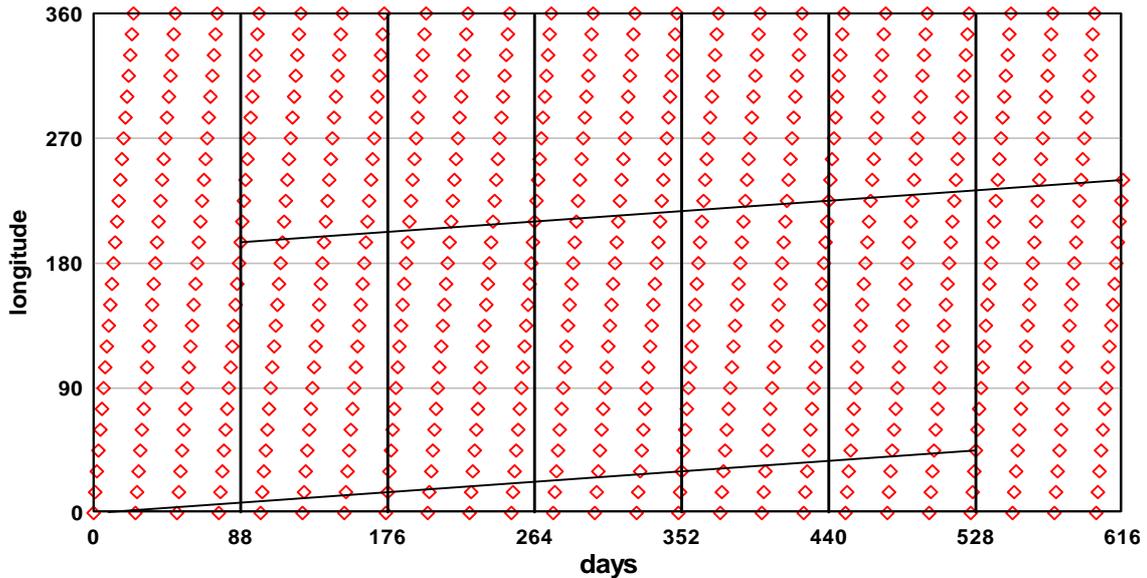

Active longitude graph.grf

**Figure 18.** Illustrates the generation of preferred longitudes for sunspot emergence. The red squares represent the daily increase in solar longitude of a point on the solar surface rotating with period 25 days. The vertical lines represent the days of closest approach to the Sun by Mercury, period 88 days, when, according to the model, sunspot emergence is more likely. The intersection points represent preferred longitudes for sunspot activity. Note that the preferred longitudes are separated by 180º in longitude.

## 5.6 Summary

We can summarize the results of this section as follows. As the time variation of the planetary model for sunspot emergence is predictable, it is possible, in most cases considered in this section, to directly compare the model time variation and the observed time variation of sunspot emergence. Because several planetary components combine in the model the unfiltered model variation is complex, as is the unfiltered time variation of sunspot emergence. Therefore, to facilitate comparison it was necessary to isolate, by band pass filtering, specific components of sunspot area emergence. Band pass filtering could have been used to isolate components in the model variation. However, it was more informative to achieve isolation by selection of the appropriate components of the model. By comparing the time variations it was possible to follow the effect of individual model pulses on sunspot emergence. The principal findings were:

(a) In most cases it was possible to observe components of sunspot area emergence lagging the stronger pulses in the model. When the model forcing was steady sunspot emergence lagged the model pulses by a few tens of days. When the model forcing changed significantly, for example by a $\pi$ phase change, sunspot emergence followed the model change over an interval of a few hundred days. The latter interval is comparable with the theoretically expected time, ~ 100 days, for the emergence of magnetic flux from the base of the convection zone to the surface, Fisher et al (2000).

(b) It was possible to confirm, for the ~ 290 day component, the observation in section 3 that the phase of this component shifts by $\pi$ from one solar cycle to the next. As discussed in section 3, this observation suggests a dependence of the phase of components of sunspot emergence on the polarity of the solar magnetic field.

(c) Some components of sunspot emergence occur in multiple episodes within solar cycles, the number of episodes varying from solar cycle to solar cycle. It appears that this is due to varying destructive and constructive interference between the several sub components within the five major groups of components indicated in Figure 6 and 7.

**6. Periodicities in sunspot area, apparently, outside the model.**

A number of peaks in the sunspot area spectrum occur at frequencies that cannot be directly associated with the model and the difference frequencies listed in Table 1. For example, Figure 19 below indicates some of the more prominent peaks in the spectra of SSAN and SSAS that do not correspond to any of the peaks in the Fourier spectrum of the model, see Figure 5 and Table 1. All of the peaks marked in Figure 19 appear prominently in the spectrum of SSAN and in the spectrum of SSAS indicating that they are significant peaks. The broad peak at about 0.0025 days$^{-1}$, not marked in Figure 19, is associated with the $f_{MV} - f_M = 0.002467$ days$^{-1}$ frequency, 405 day period, model component listed in Table 1. This periodicity has been discussed in section 3.2 and is not relevant to this section.

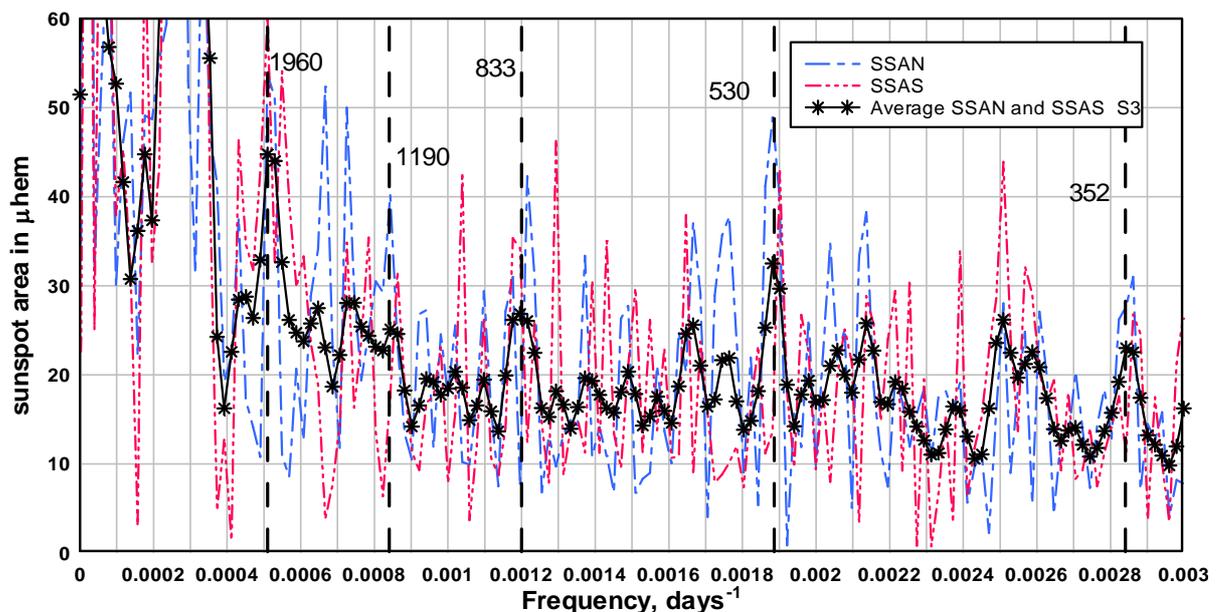

SSAN and SSAS and S3.grf
**Figure 19. The frequency spectra of SSAN and SSAS and a three point smoothed version of the average. The peaks and periods marked do not correspond to any of the peaks in the Fourier spectrum of the model indicated in Figure 5 and Table 1.**

The marked peaks are obviously Fourier components in the sunspot area records. However, as noted, there are no Fourier components at these periods in the model spectra, see Figures 5, 6 and 7. In this section we demonstrate that the periodicities marked in Figure 19 arise from sequences in the time variation of the model that repeat at varying intervals and can therefore be identified by autocorrelation.

**6.1 Repeating sequences in the model assessed by autocorrelation.**
Figure 20 illustrates the first 5000 days of the model variation from day 0 on January 01, 1876. There are repeating, similar, sequences in the variation. Three of these sequences are marked in Figure 20. The marked sequences extend over intervals of about 354 days and each sequence is separated from the next sequence by a time interval of variable length. As a result there is no Fourier component in the model spectrum associated with the repeating sequences of ~ 354 day duration.

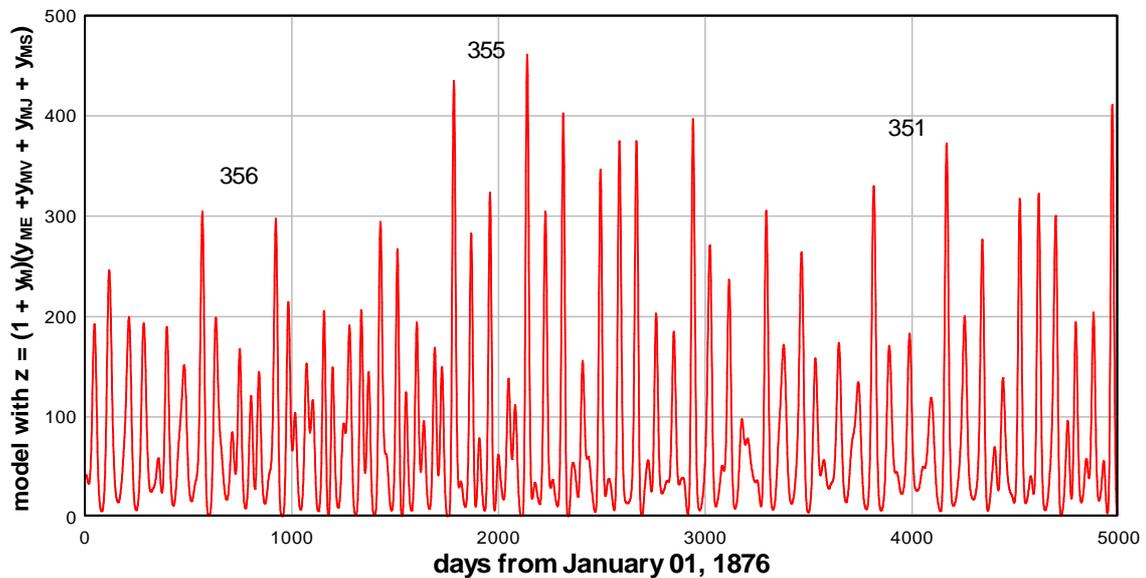

Model with reference at 250.grf

**Figure 20. Shows the first 5000 days, from day 0 on January 01 1876, of the model variation. There are repeating, similar, sequences in the variation three of which are marked. The marked sequences extend over intervals of about 354 days and each sequence is separated from the next sequence by a time interval that varies.**

An autocorrelation function provides information on repeating patterns occurring in a complex time variation. Figure 21A shows the autocorrelation function obtained over a 50,000 day interval of the model variation, $S(t) = z(t)^2$, beginning on January 01, 1876. We note that prominent autocorrelation peaks, positive peaks at 355, 530, and 2022 day delay and negative peaks at 833 and 1190 day delay, correspond, numerically, closely with the periods of the peaks indicated in the observed sunspot area spectra, Figure 19. Prominent peaks in the autocorrelation are an indication of repeated sequences in the model variation that occur more frequently than other repeated sequences.

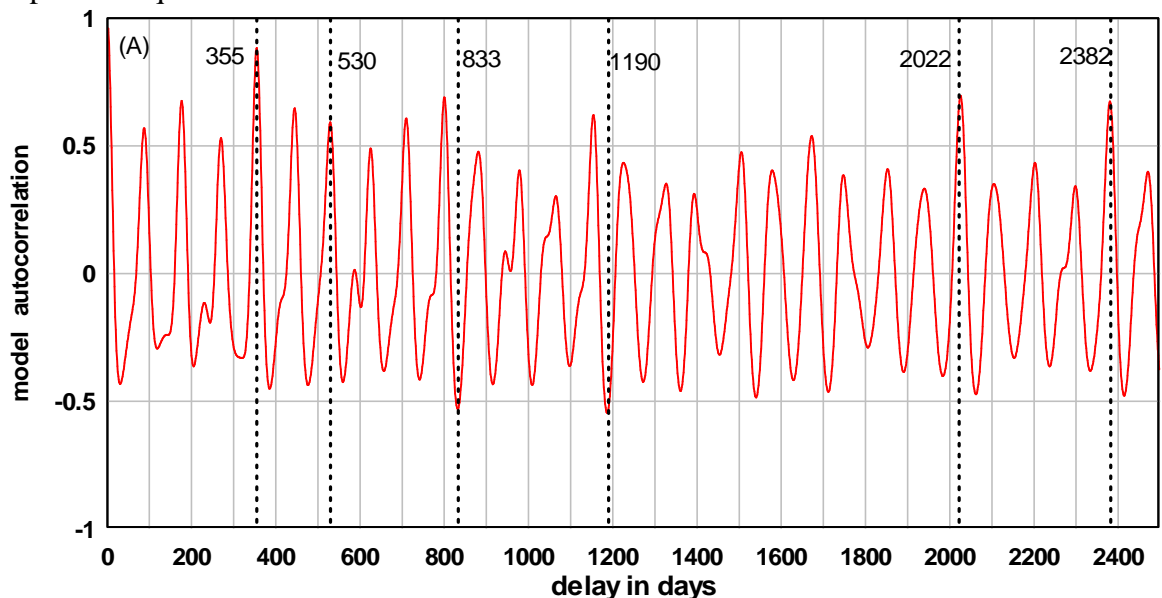

True autocorrelation model M ME MV MJ MS.grf

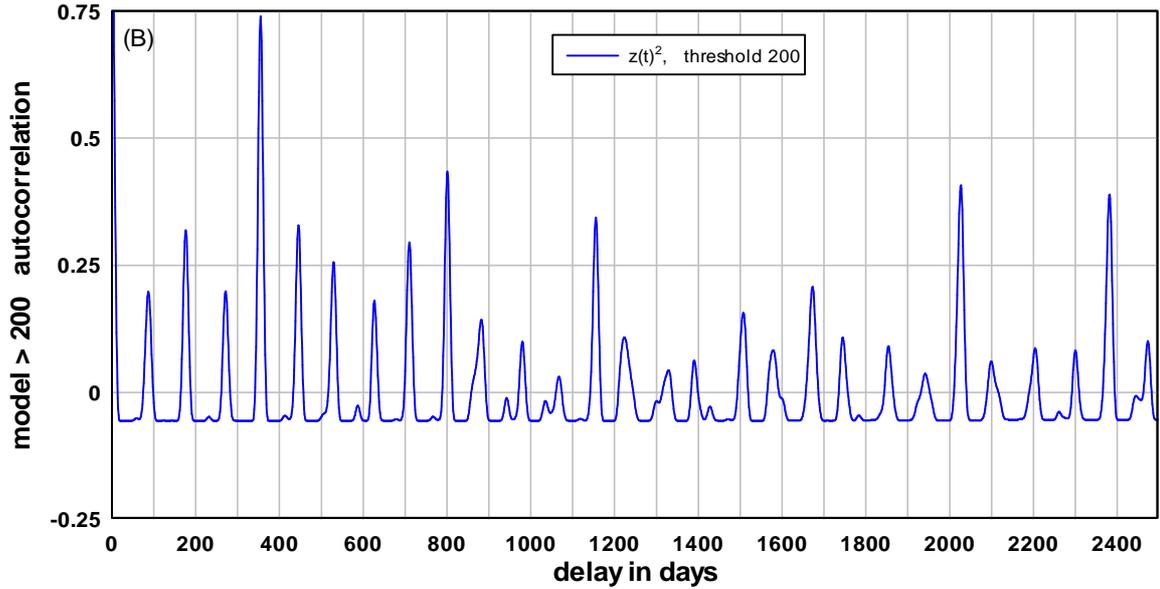

Autocorrelation zxz threshold 200.grf

**Figure 21.** (A) Shows the autocorrelation function obtained over a 50,000 day interval, beginning on January 01, 1876, of the model variation, $S(t) = z(t)^2$. Prominent autocorrelation peaks, positive peaks at 355, 530 and 2022 day delay and negative peaks at 833 and 1190 day delay, correspond numerically with the periods of the peaks in the observed sunspot area spectra, Figure 19. (B) Autocorrelation function for the model $S(t) > 200$ units.

## 6.2 Interpreting the model autocorrelation

An autocorrelation function provides information about repeating sequences in a time variation. However, it provides no information on the time when a sequence occurs. Clearly there are repeating occurrences of strong pairs of peaks in the model with separations between the peaks of ~ 355, 530, 833, 1190, 2020 and 2400 days. The numerical coincidence of the model autocorrelation delays with the periods of the spectral peaks in sunspot area spectra is evidence of an association. Figure 22 illustrates how the model autocorrelation is built up from the different contributing elements of the model. We note that the Mercury contribution, indicated with label (1+M) in Figure 22 represents the contribution due to the component proportional to the Mercury tidal variation, equation (1). This variation in the autocorrelation of Figure 22 corresponds to the presence of the Fourier component in the model at $T_M$ ~ 88 day period. The other major Fourier components in the model, at the periods of the Mercury-planet conjunctions, contribute to the model at periods of $T_{ME}$ ~ 58 days, $T_{MV}$ ~ 72 days, $T_{MJ}$ ~ 45 days and $T_{MS}$ ~ 44 days. It is evident that variations at these periodicities combine to form the overall pattern of peaks in the model time variation. This combination results in the complex model time variation and in the autocorrelation for the complete model, labeled ME MV MJ MS x (1 +M) in Figure 22. We note that if the phase angles of components of the model were set to zero and the model computed from day 0, the components would add in-phase at 355 days and at 2020 days and would add in anti-phase at 833 days and 1190 days. As an example, in-phase addition at 355 days follows from 355 ~ 8 x 44 ~ 8 x 45 ~ 6 x 58 ~ 5 x 72 ~ 4 x 88. It seems that when the model components at the true phase angles are combined it is a similar form of in-phase and out-of-phase addition that results in the prominent positive or negative autocorrelation peaks in Figure 21A and Figure 22 at ~ 355, 833, 1190 and 2020 delays. There are no Fourier components in the model $S(t) = z(t)^2$ at periods corresponding to these delays. However, when a threshold above which sunspots are triggered is applied to the model the autocorrelation function is sharpened, Figure 21B, and weak Fourier components at periods ~ 355, 833, 1190 and 2020 days appear in the model spectrum. For example setting a threshold of 200 units above which the model variation is effective results in components at ~129, 137, 240 355, 365, 455, 807, 1137, 1472 and 1668 days appearing as weak Fourier components in the frequency spectrum of the model. This suggests that a threshold level is necessary if a planetary model is to encompass the full range of mid range periodicity in sunspot area emergence.

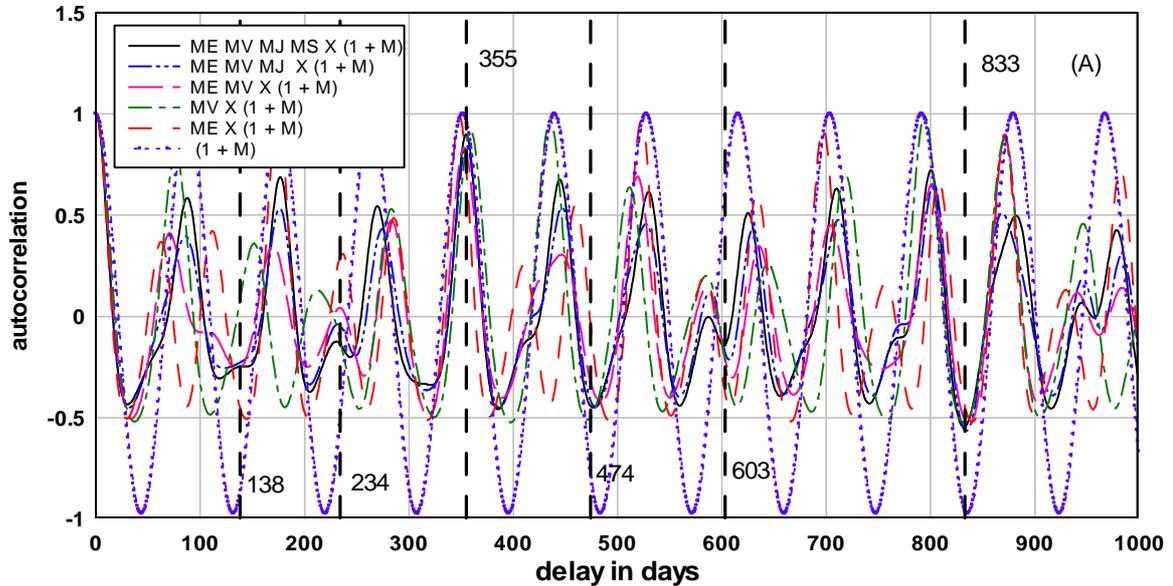

Autocorrelation all types ME MV MJ MS (1+M).grf

**Figure 22. Illustrates how the model autocorrelation is built up from different contributing elements of the model. We note that the Mercury contribution, indicated with label (1+M) in Figure 22 represents the contribution due to the component proportional to the Mercury tidal variation, equation (1).**

We note that there appears to be two types of periodicities in sunspot area emergence. One type, where Fourier components observed in sunspot emergence can be linked directly to Fourier components in the planetary model without threshold, referred to here as "direct components". A second type, where Fourier components observed in sunspot emergence can be linked to peaks in the model autocorrelation and are weakly evident in the Fourier spectrum of a model with threshold applied, referred to here as "indirect components".

**6.3 Minor peaks in the model autocorrelation associated with sunspot area periodicity.**
There are a few major peaks in the average spectrum of sunspot emergence that have not been considered previously. Four of these are indicated in Figure 23 below. The peaks at 605 days and 472 days can be identified with the autocorrelation peaks on either side of the peak at 530 day delay in Figures 21 and 22. However, we are interested in the strong spectral peak at period 235 days, frequency 0.00425 days$^{-1}$ and the strong spectral peak at period 138 days, frequency 0.00725 days$^{-1}$ in Figure 23. Neither of these periodicities is near any of the Fourier components of the model. However, weak peaks in the model autocorrelation function, Figure 21 and Figure 22, appear to be associated with these peaks in the sunspot area spectrum and, as mentioned above, weak Fourier components at ~ 137 days and ~ 240 days occur as Fourier components in a model with threshold applied. The autocorrelation peaks referred to are indicated in the autocorrelation of Figure 22, estimated to be at 138 days delay and 234 days delay. We therefore assign the 138 day and 235 day peaks in the sunspot area spectrum as belonging to the group of indirect periodicities. It can be ascertained from Figure 22 that the 138, 234 and 603 day peaks appear to be associated with the ME(1+M) autocorrelation. The peak at 474 days appears to be associated with the MV(1+M) autocorrelation. It appears that the amplitude of positive and negative peaks in the model autocorrelation does not necessarily correspond to the amplitude of peaks observed in the spectra of sunspot emergence. The indirect periodicities associated with the model autocorrelation discussed in this section now include the ~ 138, 235, 355, 472, 530, 605, 833, 1190 and 2020 day periodicities observed in the sunspot area record. Direct periodicities are associated with the ten frequency difference components listed in Table 1.

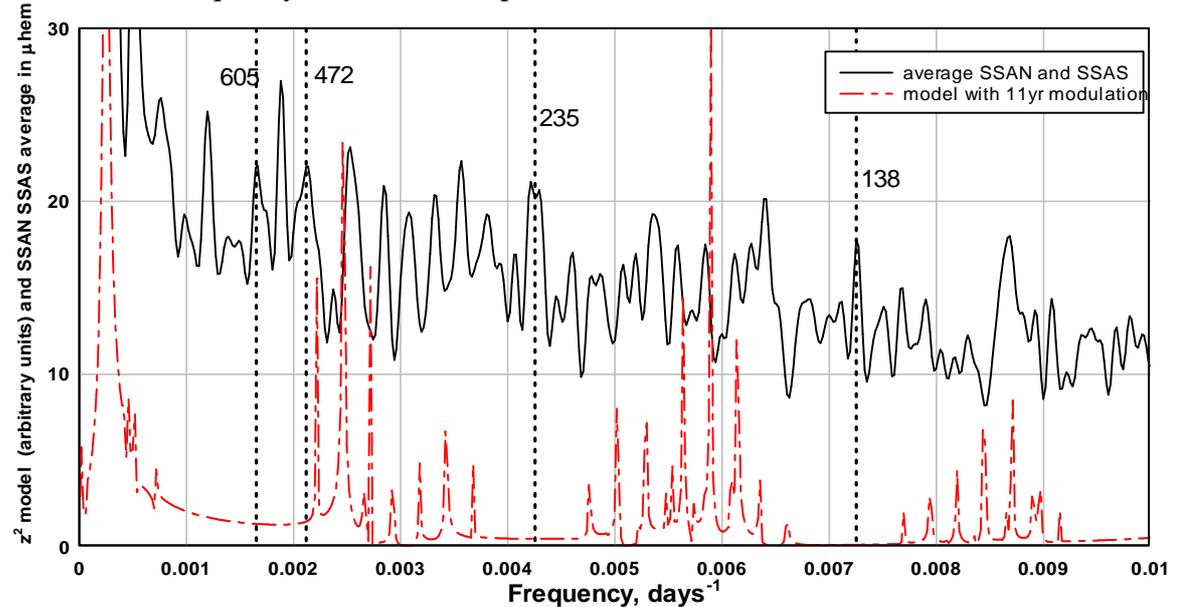

Y2 model shows 425 and 138.grf

**Figure 23. Illustrates the occurrence of strong peaks in the observed sunspot area spectrum that are not associated with Fourier components of the model but are associated with less prominent peaks in the model autocorrelation.**

**6.4 Comparing direct and indirect periodicities in sunspot emergence.**
There are distinct differences in behaviour between direct and indirect components of sunspot area emergence. The principal difference is that it is possible, as illustrated in section 4, to use the planetary model to predict the time variation of direct components of sunspot emergence. For example, the peaks in the ~ 176 day component can be shown to occur at ~ 50 day lag to the model peaks when the model variation is stable, Figure 12. However, it is difficult to determine, from the planetary model, the time dependence of the indirect components of sunspot emergence.

A second significant difference is that the spectral peaks of the indirect components of sunspot emergence usually occur as narrow peaks in the sunspot area spectra, e.g. the narrow peaks at 138, 235, 355, 530, or 833 day period. Whereas, spectral peaks associated with direct components of sunspot area tend to occur in groups of peaks or as a broad peak centred on one of the Fourier components of the model, e.g. peaks associated with the 116, 292 or 405 day periods. The reason for this appears to be that direct components of sunspot emergence are strongly modulated by the ~ 11 year cycle. The modulation generates strong sidebands resulting in multiple peaks or a broad peak in the spectrum. For example, the ~290 day component, the time variation of which, Figure 9, illustrates strong solar cycle modulation over the entire record, gives rise to four near equal amplitude peaks in the spectrum, Figure 7. On the other hand, as we show below in Figure 24, the time variations of indirect components of sunspot area are not strongly modulated by the ~ 11 year cycle. As a result there appear to be no significant modulation sidebands and strong narrow peaks occur in the average sunspot area spectrum. As an example, Figure 24 illustrates the time dependence of the ~352 day and the ~830 day components of sunspot area North. The ~ 11 year solar cycle of sunspot area emergence is also indicated. Clearly the modulation of the ~ 352 and ~ 830 day components is not strongly associated with the ~11 year solar cycle and the modulation that does occur, most noticeably a π radian phase change during solar cycle 19, at day 30000, does not lead to significant broadening of the spectral peaks.

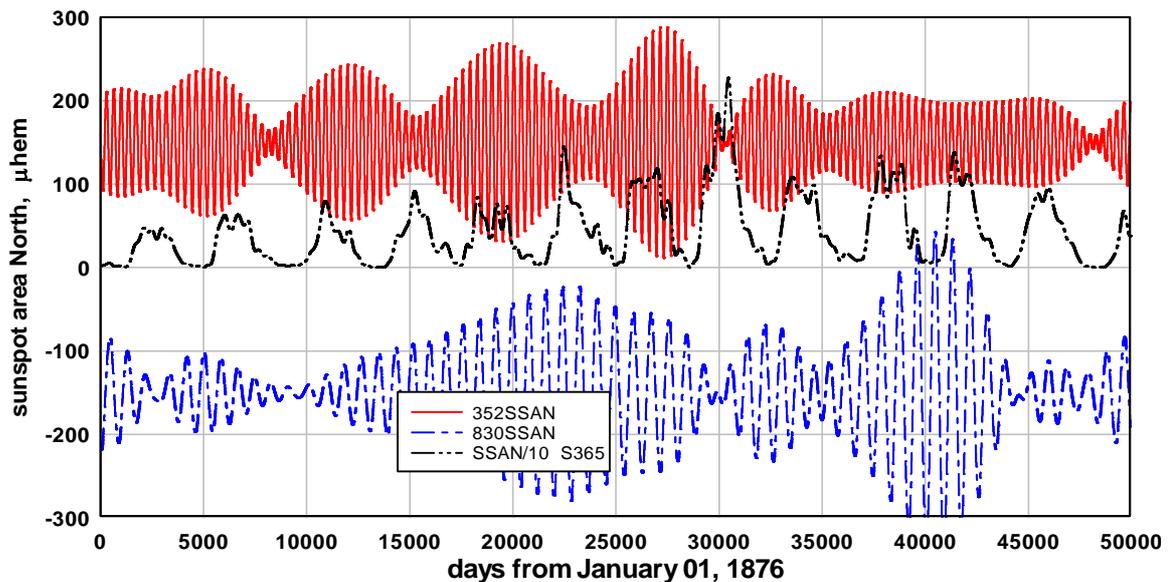

Compares 352SSAN and 830SSAN.grf

**Figure 24.** Shows the time dependence of the ~352 day and the ~830 day components of sunspot area North. The ~ 11 year solar cycle of sunspot area emergence is also indicated. Clearly the ~352 and ~830 day components are not strongly amplitude modulated by the ~11 year solar cycle. The strongest modulation common to both variations, a π phase change during solar cycle 19, occurs near day 30000 and is probably associated with the solar magnetic field reversal during solar cycle 19.

As indirect components are not strongly amplitude modulated by the solar cycle they tend to contribute disproportionately to the variation of sunspot area emergence during solar minima. For example, referring to Figure 24 above, it is expected that the ~352 day and ~830 day components would contribute strongly to the small level of sunspot area emergence in the solar cycle minimum between solar cycles 21 and 22, ~ day 40000.

**6.5  Effect of solar magnetic field reversal.**
As discussed in section 3.2 the phase of a component of sunspot area emergence appears to depend on the sign of the solar magnetic field. In Figure 24 above it is evident that sharp amplitude decreases accompanied by phase reversals of the ~352 and ~830 day components occasionally occur near the times of solar magnetic field reversal around solar cycle maximum. For example, in Figure 24, at the maximum of solar cycle 19, both the ~352 day and ~830 day components suffer a sharp minimum in component amplitude and a π phase change in component variation. Another example in Figure 24 occurs near the maximum of solar cycle 12. This effect is also noticeable in direct components. The ~ 405 day direct component of SSAN and SSAS show this effect in several solar cycles. We illustrate with two cases in Figure 25. The effect is evident in solar cycle 18 when, near day 26050, the 405SSAN component suffers a $\pi$ phase change, Figure 25A, and in solar cycle 21 when both of the 405SSAN and 405SSAS components suffer a $\pi$ phase change near day 37835, Figure 25B. To facilitate following the component phase changes the time axes are divided into 405 day intervals.

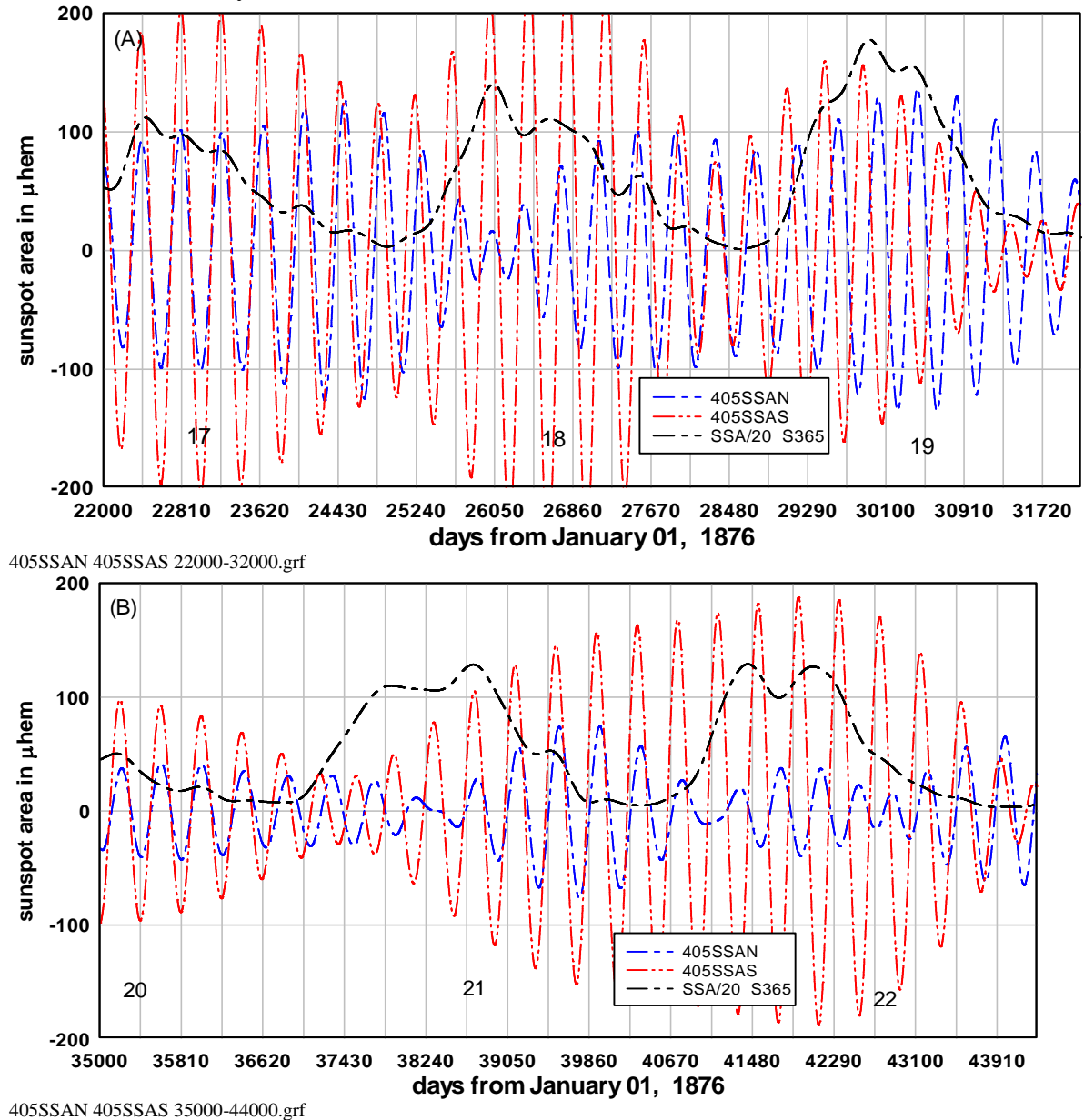

**Figure 25. (A) In solar cycle 18 the 405SSAN component suffers a $\pi$ phase change while the 405SSAS component remains at constant phase. (B) In solar cycle 21 both the 405SSAN and 405SSAS components suffer a $\pi$ phase change. To facilitate following the component phase change the time axes are divided into 405 day intervals.**

As we have seen in section 3.2, $\pi$ radian phase changes associated with magnetic field reversal have the effect of splitting the spectral lines in the spectra of sunspot area. Clearly the phase changes have a strong effect on the time dependence of sunspot emergence with the decrease in amplitude of a component in response to the phase change leading to sharply divided episodes of sunspot emergence during individual solar cycles. Figure 26 illustrates an example where solar magnetic field reversal may play a role. As discussed in section 5.2 we expect the model with $z(t) = [1 + y_M(t)][y_{ME}(t) + y_{MJ}(t)]$ to provide for multiple episodes of the 176SSAN component in the earlier solar cycles like solar cycle 13. However, the model does not allow for phase changes due to solar field reversal. In Figure 26 $\pi$ phase changes occurring in the model variation are marked with dotted lines. The broken blue line is the variation of component 176SSAN during solar cycle 13. During the first part of the cycle, day 5000 to the first phase change, sunspot emergence responds to model forcing with ~ 30 day lag. After the first phase change 176SSAN is out of phase and begins to decrease. After the second phase change 176SSAN is in-phase and lagging the stronger model peaks by ~ 50 days but continues to decrease even as the model peaks increase. As the model peaks reach maximum strength, 176SSAN reaches a minimum and suffers a $\pi$ phase change. As this phase change in 176SSAN does not match a model phase change it is assigned as a phase change and minimum in 176SSAN associated with solar magnetic field reversal occurring around the middle of the solar cycle. It is difficult to verify this as the timing of solar magnetic field reversal has only been measured in recent times.

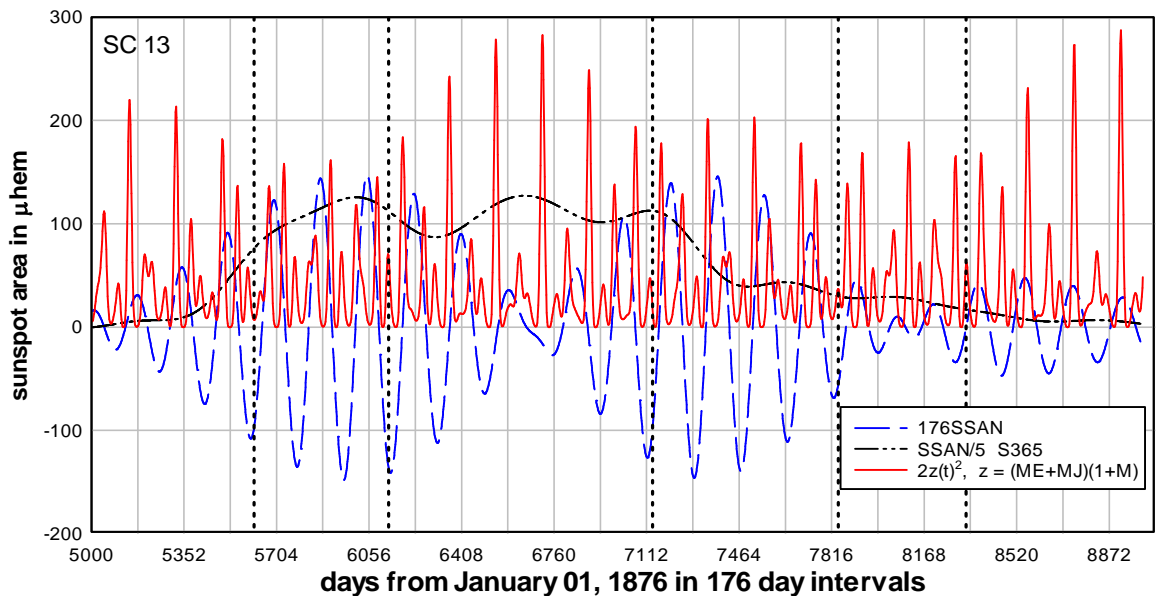

Compares model (ME+MJ)(1+M) with 176 SSAN SC13.grf

**Figure 26. The full red line is the model, $S(t) = 2z(t)^2$ with $z(t) = [1 + y_M(t)][y_{ME}(t)+y_{MJ}(t)]$. The $\pi$ phase changes in the model are marked with dotted lines. The broken blue line is the variation of component 176SSAN during solar cycle 13. During the first part of the cycle, day 5000 to the first phase change, sunspot emergence responds to model forcing with ~ 30 day lag. After the first phase change 176SSAN is out of phase and begins to decrease. After the second phase change 176SSAN is lagging by ~ 50 days but continues to decrease as the model peak increases. As the model peaks reach maximum, 176SSAN reaches a minimum and suffers a $\pi$ phase change. As this phase change does not match a model phase change it is assigned as due to solar magnetic field reversal occurring around the middle of the solar cycle.**

## 8. Correlation between the North and South components of sunspot emergence.

The observed time dependence of sunspot area emergence and the corresponding frequency spectrum are both complex. Thus any model of sunspot emergence must generate a complex time variation to match observations. Much of the prior work on the ~11 year periodicity and the intermediate range periodicity in sunspot records assumes the complexity arises from either periodic processes in the Sun, Charbonneau (2002), or random processes on the Sun, Wang and Sheeley (2003). Evidence against the source being a random process is that, in the frequency spectra derived from sunspot area on the North and the South hemispheres of the Sun, prominent low frequency peaks in both spectrums occur at or close to the same frequency, Figure 1 and Figure 19. Similar North and South spectrums would be unlikely if the periodicity arose from random sunspot emergence on the Northern hemisphere and random sunspot emergence on the Southern hemisphere. However, similar spectra from both hemispheres would be expected if a systematic effect influenced sunspot emergence on both hemispheres. In the latter case we would expect the time variations in sunspot area on each hemisphere to be correlated.

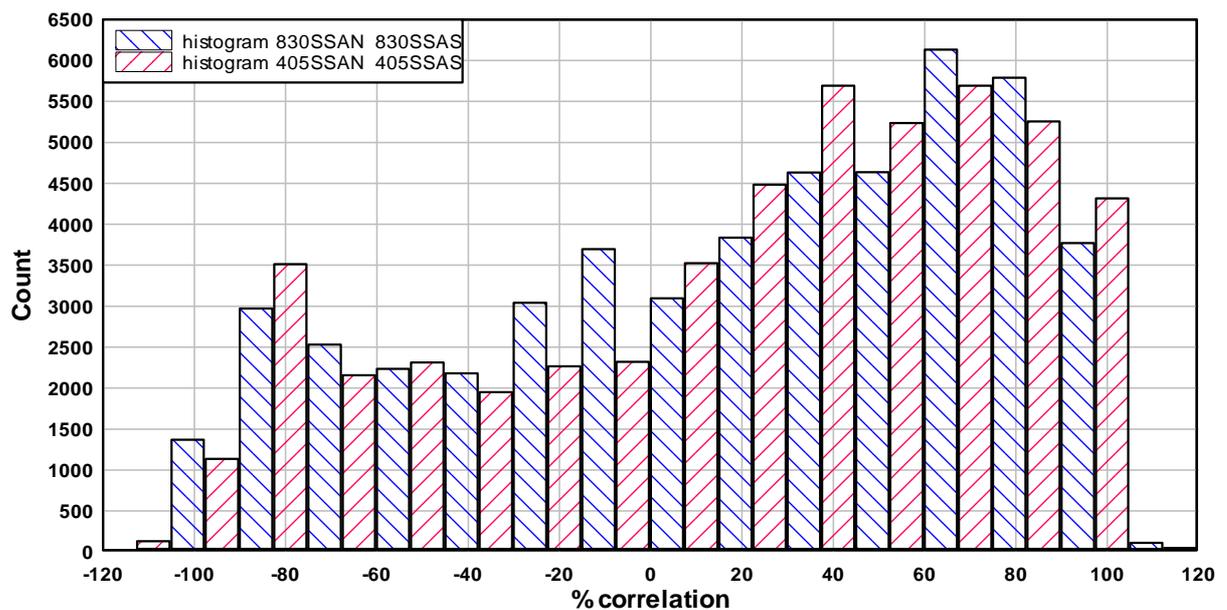

Histogram 830 and 405.grf

**Figure 27. Histograms of the percentage correlation of the ~405 day North and South components and the percentage correlation of the ~830 day North and South components of sunspot area. The average positive correlation coefficient, about +0.26, indicates the tendency for the North and South components of sunspot emergence to be positively correlated for these components.**

When the correlation between SSAN and SSAS components is estimated using the method outlined in section 4.1 and a histogram of the percentage correlation coefficient is made we get an indication of the distribution of correlation between the North and South components. Figure 27 shows the histograms for the percentage correlation of the ~405 day North and South components and the percentage correlation of the ~830 day North and South components. This shows the tendency for some North and South components to be positively correlated for much of the record suggesting, for these components, a common origin influencing both the North and South sunspot emergence. The average correlation coefficients are +0.27 for the ~405 day components and +0.25 for the ~830 day components. Other components do not show the same positive correlation so the observations, Figure 27, may not be significant. It is outside the scope of this paper to study this aspect further.

**9. Conclusion.**

The time variation and the frequency spectrum associated with sunspot area emergence are complex. However, sunspot area data has the advantage of providing two data records, one each from the North and South hemispheres of the Sun. By comparing spectra from the North and South records it was possible to identify spectral peaks common to both records, adding significance to those peaks. Based on the intermediate frequency components observed a simple empirical planetary model of sunspot emergence based on Mercury – planet conjunctions was developed. A non-linear relation connecting sunspot emergence and the effect of conjunction was introduced into the model and this generated frequency difference terms in the model spectrum that coincided, broadly, with the frequencies of spectral peaks identified in sunspot area spectra. By adding the effect of solar cycle amplitude modulation and solar magnetic field reversal every solar cycle it was possible to use the model to predict fine detail in the spectra and the time dependence of sunspot area emergence. The principal findings were:

The basic frequency components in the intermediate frequency range sunspot area spectrum are predominantly due to components at frequency differences derived from the Mercury-planet conjunction frequencies and the frequency of Mercury.

Amplitude modulation of sunspot area emergence by the ~ 11 year solar cycle results in strong sidebands at either side of the basic frequency components. Phase modulation of sunspot area components associated with solar magnetic field reversal during each solar cycle further splits the basic frequencies in the spectrum. By including a phase modulation term it was possible to use the planetary model to predict the splitting of peaks in parts of the spectrum precisely.

In individual solar cycles it is possible to observe the close relationship between the time dependence of components of sunspot area emergence and the model time variation. In particular, when the model time variation is stable sunspot area emergence lags the model variation by a few tens of days. When the model variation changes by $\pi$ radians the component of sunspot area emergence follows the model change by decreasing in amplitude to near zero and re-emerging with a $\pi$ phase change to be in-phase with the model variation. These major changes in phase and amplitude of sunspot area emergence occur over intervals of a few hundred days and have the effect of separating sunspot area emergence during a solar cycle into episodes of variable length ranging from about two years upwards. That there is, in many cases, a good match between time variation of sunspot area emergence and the time variation of the planetary model suggests the model is representative of an actual planetary forcing of sunspot emergence.

The observational evidence indicates that phase reversals in sunspot area emergence also occur on solar magnetic field reversal. This effect was approximately included in the model so that spectral features of sunspot emergence could be followed. However, it is yet to be included in the model to facilitate following changes in the time variation of sunspot emergence associated with magnetic field reversal. In a more complete model it may be possible to follow model phase changes and magnetic field reversal phase changes over the entire record and it may be possible to observe interference between the two effects when the effects overlap in time.

There are components of sunspot area emergence, corresponding to prominent peaks in sunspot area spectra, which cannot be identified with Fourier components in the model spectrum but can be identified with peak delays in the autocorrelation function of the model. Some of these components can be identified in a model that includes a threshold for triggering sunspot emergence. With the inclusion of these components most of the intermediate range spectral peaks in sunspot area spectra can be associated with the planetary model either directly via Fourier components or indirectly via autocorrelation.

Finally, while the model is simple, empirical and incomplete, the fact that it predicts the above several features of sunspot emergence provides an observational constraint on any physical mechanism relating to intermediate range periodicity in sunspot area emergence.

**References.**